 \documentclass[12pt]{elsarticle}

\usepackage{amssymb}
\usepackage{balance}

\usepackage[usenames]{color}
\newcommand{\cblue}[1]{ }
\newcommand{\cred}[1]{ }

\newcommand{\bk}{{\bf k}}

\newcommand{\rr}{{\bf r}}

\usepackage{lipsum}
\makeatletter
\def\ps@pprintTitle{%
 \let\@oddhead\@empty
 \let\@evenhead\@empty
 \def\@oddfoot{}%
 \let\@evenfoot\@oddfoot}
\makeatother

\begin{document}

\begin{frontmatter}

\title{Transparent boundary conditions for time-dependent electron transport in the R-matrix method with
applications to nanostructured interfaces}

\author[1,2]{G. A. Nemnes\corref{mycorrespondingauthor}}

\author[1]{Alexandra Palici}

\author[3]{A. Manolescu}

\cortext[mycorrespondingauthor]{Corresponding author. Tel.: +40 (0)21 457 4949/157. \\ {\it E-mail address:} nemnes@solid.fizica.unibuc.ro (G.A. Nemnes).}
\address[1]{University of Bucharest, Faculty of Physics, Materials and Devices for Electronics and Optoelectronics Research Center,\\ 077125 Magurele-Ilfov, Romania}
\address[2]{Horia Hulubei National Institute for Physics and Nuclear Engineering, 077126 Magurele-Ilfov, Romania}
\address[3]{School of Science and Engineering, Reykjavik University, Menntavegur 1, IS-101 Reykjavik, Iceland}




\begin{abstract}
Transparent boundary conditions for the time-dependent Schr\"odinger equation are implemented
using the R-matrix method. The employed scattering formalism is suitable for describing 
open quantum systems and provides the framework for the time-dependent coherent transport.
Transmission and reflection of wave functions at the edges of a finite quantum system are essential for an
accurate and efficient description of the time-dependent processes on large time scales.
We detail the computational method and point out the numerical advantages stemming from the 
open system approach based on the R-matrix formalism.
The approach is used here to describe time-dependent transport across nanostructured interfaces 
relevant for photovoltaic applications. 
\end{abstract}

\begin{keyword}
transparent boundary conditions \sep electron transport \sep
time-dependent Schr\"odinger equation \sep scattering formalism \sep nanostructured interface
\end{keyword}

\end{frontmatter}

%\linenumbers

\section{Introduction}

By continuing miniaturization the current electronic devices have already reached length scales of only 
a few tens of nanometers and, in the past few years, quantum mechanical approaches have been extensively used 
 for the modelling of the electron transport down to molecular scale \cite{tao,popescu}.  
In particular, there is a lot of interest for an efficient description of time-dependent coherent transport
and examples of physical systems may include
nanoscopic antennas \cite{gennaro}, 
electron and hole transport through nanostructured interfaces with applications
in photovoltaics \cite{loi}, high frequency transistors \cite{pallechi}, coherent phonon pulses 
in the description of transient thermal transport \cite{chen} etc.

Time-dependent charge transport has been investigated in a number of studies, using different techniques. 
The Green-Keldysh formalism has been applied to transport in mesoscopic systems having external 
time-dependent voltages \cite{jauho} or barriers \cite{moldoveanu1}. 
\cblue{The generalized master equation (GME) has been used for charge transport via many-body electron systems \cite{moldoveanu2,moldoveanu3} or electron-photon complex states \cite{vidar,nzar}.}
The Lippmann-Schwinger equation was employed in the context of wave packets propagation in quantum wires with
 magnetic fields \cite{thorgilsson}.
More recently the time-dependent wavepacket diffusion (TDWPD) method was employed as an
approximation to the exact stochastic Schr\"odinger equation (SSE) method \cite{zhong}.

In the description of the time-dependent evolution of the wavefunctions an essential ingredient is
represented by the {\it transparent boundary conditions} (TBCs) \cite{antoine}. 
Similar problematics is found in beam propagation in optics \cite{hadley}.
As the system under direct numerical 
investigation is finite, the TBCs are required in order to ensure the wave propagation over the boundaries. 
Otherwise, the waves may be partly or completely reflected back into the region of interest, bringing a limitation regarding
the maximum time scales for which the process can be investigated. These are related to the size of
the scattering region, which, in turn, determines the computational cost.  

In this context we employ the R-matrix method as an efficient approach to obtain the stationary scattering
functions. The formalism was developed by Wigner and Eisenbud \cite{wigner} in the field of nuclear physics and 
later was employed to obtain the transport properties of mesoscopic devices \cite{smrcka,wulf,onac}. It has been further applied to describe
coherent charge transport in nanotransistors \cite{nemnes1,nemnes2,nemnes3}, 
thermopower in quantum wires \cite{nemnes4}
spin dependent transport \cite{nemnes5} and to investigate the effects of graded distribution of scattering centers on ballistic transport \cite{nemnes6}
and charge localization in dendritic interfaces relevant for photovoltaic applications \cite{nemnes7}. \cblue{Most recently, the R-matrix method was employed describing snaking states in core-shell nanowires \cite{manoles1}, their signature being experimentally validated in a quantum point contact register \cite{heedt}.}
Here we use the advantages of this approach as a basis for solving the time-dependent problem   
with TBCs.

Furthermore, an accurate description of many electronic devices of practical interest usually
require a quantum approach at small length scales (e.g. atomistic, effective mass models)
and a classical description at macroscopic scales (e.g. drift-diffusion type models), 
usually combined in hybrid transport models, such as the drift diffusion model (QDD) \cite{degond}, 
the quantum corrected drift diffusion model (QCDD) \cite{falco} or the Schr\"odinger-Poisson-Drift-Diffusion model (SPDD) \cite{pirovano}.
In this context of transport model hierarchies we discuss the possibility of using the developed time-dependent framework 
for a local characterization of the photo-current and charge separation that occurs in the vicinity of the
nanostructured interfaces of photovoltaic devices.  

The paper is organized as follows. In the next section the generic model system is indicated and 
the general R-matrix formalism is presented, pointing out the efficiency of the method in constructing the stationary scattering wavefunctions for a relatively large set of total energies.
The time dependent solutions are then determined using the eigenvectors of the open quantum system.
The computational method is discussed in detail.
In the following section the charge transport across nanostructured interfaces relevant for 
photovoltaic applications is analyzed. This includes the time-dependent description of the charge separation
and photo-current near the interface.

\section{Model and Computational method}

\subsection{Coherent scattering model}

\begin{figure}[h]
\centering
\includegraphics[width=6.cm]{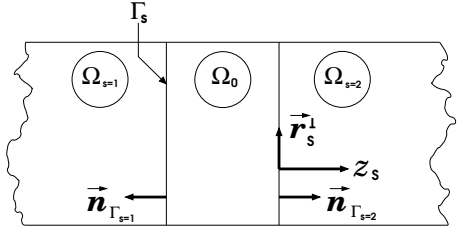}
\caption{Two-terminal scattering model system: scattering region $\Omega_0$ connected by leads, represented
by $\Omega_s$ domains. The interfaces between $\Omega_0$ and $\Omega_s$ are denoted by $\Gamma_s$.
A reference system is indicated in each lead $(z_s,\rr_s^\perp)$.}
\label{scat_model}
\end{figure}

The model system, depicted in Fig.\ \ref{scat_model}, consists of a central finite region, 
$\Omega_0$, connected by semi-infinite leads, $\Omega_s$, which describes a multi-terminal device.
This framework is typically used in the Landauer-B\"uttiker formalism and can also employed in atomistic transport calculations \cite{bell}. Carriers are injected from each terminal and elastically scattered in the central region $\Omega_0$.
%while we denote the entire system by $\Omega=\Omega_0\bigcup_s\Omega_s$.
The potential energy $V(\rr\in\Omega_0)\equiv V_{\Omega_0}$ in central region is arbitrary,
while the potentials along the lead direction in the $\Omega_s$ domains
are constant,
i.e. $V(\rr\in\Omega_s)\equiv V_{\Omega_s}=V^\perp(\rr^\perp_s)+V^\parallel_s$.

\subsection{The R-matrix method}

We consider the stationary Schr\"odinger equation 
\begin{equation}
\label{sch}
{\mathcal H}\Psi(\vec{r}) = E\Psi(\vec{r}),
\end{equation}  
with ${\mathcal H}=-\hbar^2/(2m^*)\bigtriangleup + V(\rr)$,
subject to scattering boundary conditions, i.e. the particles are coming from
one of the leads $\Omega_s$, commonly termed as asymptotic boundary conditions,
since the electrons are coming from $z_s\rightarrow\infty$. The solutions of Eq.\ (\ref{sch})
are the scattering functions $\Psi(\vec{r})$ corresponding to the total energy $E$, 
which may vary continuously.

The wavefunctions inside the leads can be generically written as:
\begin{eqnarray}
\Psi_s(\vec{r} \in \Omega_s;E)
&=& \sum_i  \Psi_{\nu}^{in} \exp{(-ik_{\nu} z_s)}
     \Phi_\nu(\rr^\perp_s) \nonumber\\
    &+& \sum_i \Psi_{\nu}^{out} \exp{(ik_{\nu} z_s)}
        \Phi_\nu(\rr^\perp_s), 
\end{eqnarray}
where $k_{\nu} = \sqrt{ {2m^* \over \hbar^2}(E-E_{\perp}^{\nu}) }$ are the wavevectors
along the transport direction in each channel $\nu$.
The composite index $\nu=(s,i)$ denotes the channel $i$ from lead $s$.
The complex coefficients $\Psi_{\nu}^{in}$ and $\Psi_{\nu}^{out}$ are the amplitudes of the 
incoming and outgoing wavefunctions for each channel, with respect to the scattering region $\Omega_0$.
The energies $E_{\perp}^{\nu}$ correspond to the transverse modes 
$\Phi_\nu$ inside the leads, which are solutions of the transversal Schr\"odinger equation 
$(\rr^\perp_s\in\Omega_s)$:
\begin{equation}
\label{sch_perp}
\left[-\frac{\hbar^2}{2m^*}\bigtriangleup_{\perp;s} + V(\rr\in\Omega_s)\right] \Phi_\nu(\rr^\perp_s) = E_{\perp}^{\nu} \Phi_\nu(\rr^\perp_s).
\end{equation}  

To obtain the solution of the scattering problem, one solves first the auxiliary Wigner-Eisenbud (W-E) 
problem, which is defined on the scattering region $\Omega_0$ as the same Schr\"odinger equation as in Eq. (\ref{sch}), with new boundary conditions on the interfaces $\Gamma_s$:     
\begin{eqnarray}
\label{we1}
& &{\mathcal H} \chi_l(\rr\in\Omega_0) = E_l \chi_l(\rr\in\Omega_0),\\
\label{we2}
& &\;\;\;\left[ \frac{\partial \chi_{l}}{\partial z_s} \right]_{\Gamma_s} = 0.
\end{eqnarray}  
One may further write the scattering functions in $\Omega_0$ as linear combinations of the W-E eigenfunctions $\chi_l$ as:
\begin{equation}
\Psi(\rr\in\Omega_0 ; E) = \sum_l a_l(E) \chi_l(\rr\in\Omega_0).
\end{equation}
By imposing the continuity relations on each lead - scattering region interface $\Gamma_s$,
for the scattering functions and the current:
\begin{eqnarray}
\label{contcond1}
\left. \Psi(\rr) \right|_{\Gamma_s} &=& \left. \Psi_s(\rr) \right|_{\Gamma_s} \\
\label{contcond2}
\left. \frac{1}{m^*} \frac{\partial \Psi(\rr)}{\partial z_s} \right|_{\Gamma_s}
&=& \left. \frac{1}{m^*} \frac{\partial \Psi_s(\rr)}{\partial z_s} \right|_{\Gamma_s}
\end{eqnarray}
one obtains the relation between the incoming and outgoing
coefficients $\Psi_{\nu}^{in}$ and $\Psi_{\nu}^{out}$.
In a compact form it may be written as $\vec{\Psi}^{out} = {\rm S} \vec{\Psi}^{in}$, where ${\rm S}$ is the
scattering matrix. 

The S-matrix can be conveniently expressed in terms of an R-matrix \cite{nemnes1}:
\begin{equation}
\label{smatrix}
{\rm S} = - \left[ {\rm 1} - \frac{i}{m^*} {\rm R} {\rm k} \right]^{-1} \left[ {\rm 1} + \frac{i}{m^*} {\rm R} {\rm k} \right]
\end{equation}
where 
\begin{equation}
{\rm R}_{\nu\nu'}(E) = -\frac{\hbar^2}{2} \sum_{l=0}^{\infty} \frac{(\chi_l)_\nu(\chi_l^*)_{\nu'}}{E-E_l},
\end{equation}
with
\begin{equation}
\label{chilnu}
(\chi_l)_\nu = \int_{\Gamma_s} d \Gamma_s \Phi_\nu(\rr^\perp_s) \chi_l(\rr\in\Gamma_s).
\end{equation}
The ${\rm k}$-matrix is diagonal, ${\rm k}_{\nu\nu'}=k_\nu\delta_{\nu\nu'}$.

The scattering functions can be determined for each energy using the scattering S-matrix.
Assuming the particle is incident on channel $\nu$, i.e. from lead $s$ and having the transversal mode $i$, 
one may write the wavefunction inside the lead $s''$ as:
\begin{eqnarray}
\Psi_\nu (\rr\in\Omega_{s''} ; E) &=& \frac{1}{\sqrt{2\pi}} \sum_{\nu'} \big[ \exp{(-ik_{\nu'} z_{s'})} \; 
						\delta_{\nu'\nu}  \nonumber\\
                          &+&  {\rm S}^t_{\nu\nu'} \exp{(ik_{\nu'} z_{s'})} \big] 
						\Phi_{\nu'}(\rr^\perp_{s'}) \; \delta_{s's''}
\end{eqnarray}
while in the scattering region we have:
\begin{equation}
\Psi_\nu (\rr\in\Omega_0 ; E) = 
\frac{i}{\sqrt{2\pi}}\sum_{\nu'} ({\rm 1} - {\rm S}^t)_{\nu\nu'} {\rm k}_{\nu'} \bar{R}_{\nu'}(\rr\in\Omega_0; E)
\end{equation}
with
\begin{equation}
\bar{R}_{\nu}(\rr\in\Omega_0 ; E) = \int_{\Gamma_s} d \Gamma_s \; 
      {\mathcal R}(\rr\in\Omega_0,\rr'\in\Gamma_s ; E) \; \Phi_{\nu}(\rr^\perp_{s})
\end{equation}
and
\begin{equation}
{\mathcal R}(\rr\in\Omega_0, \rr'\in\Omega_0; E) = \frac{\hbar^2}{2} \sum_l 
	\frac{\chi_l(\rr\in\Omega_0)\chi_l(\rr'\in\Omega_0)}{E-E_l}
\end{equation}

Depending on the total energy $E$ we distinguish between open (propagating) channels with real $k_\nu$
for $E_{\perp}^{\nu}\le E$, and closed (non-propagating) channels with imaginary $k_\nu$ for $E_{\perp}^{\nu}>E$.

The transmission functions can be determined for each pair of propagating modes and each energy $E$ 
from the unitary matrix 
$\tilde{\rm S}={\rm k}^{1/2}{\rm S}{\rm k}^{-1/2}$: 
${\mathcal T}_{\nu\nu'} = |\tilde{\rm S}_{\nu\nu'}|^2$. 
The total lead-to-lead transmission can be calculated as:
$T_{ss'} = \sum_{i,i'} {\mathcal T}_{\nu\nu'}$,
where the summation is performed only over the open channels.

\subsection{Time-dependent problem}

\begin{figure}[t]
\centering
\includegraphics[width=4.cm]{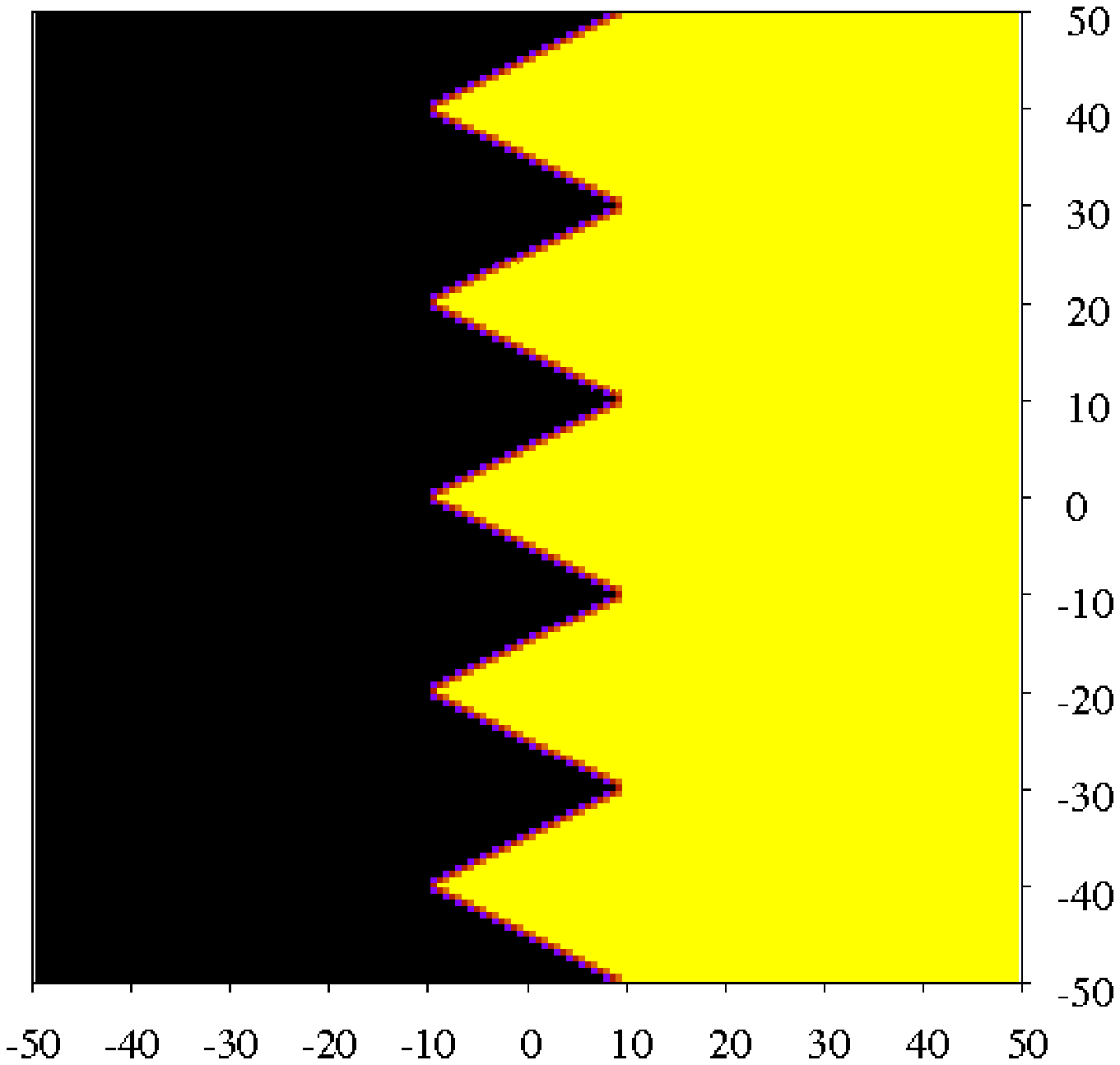}\hspace*{0.5cm}
\includegraphics[width=4.cm]{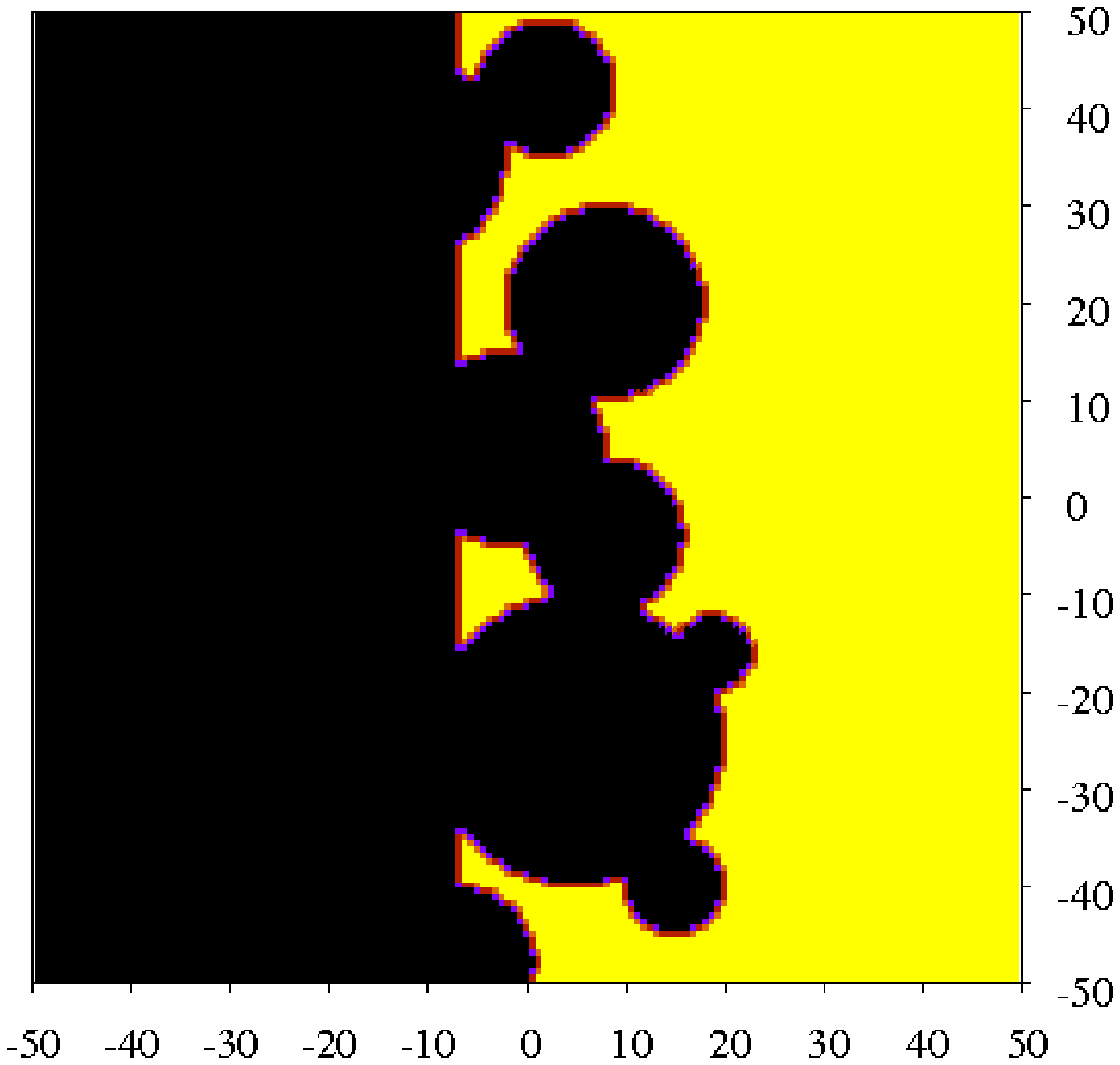}\\
\hspace*{0cm}(a)\hspace*{4.0cm}(b)\\
\vspace*{0.3cm}\includegraphics[width=7.cm]{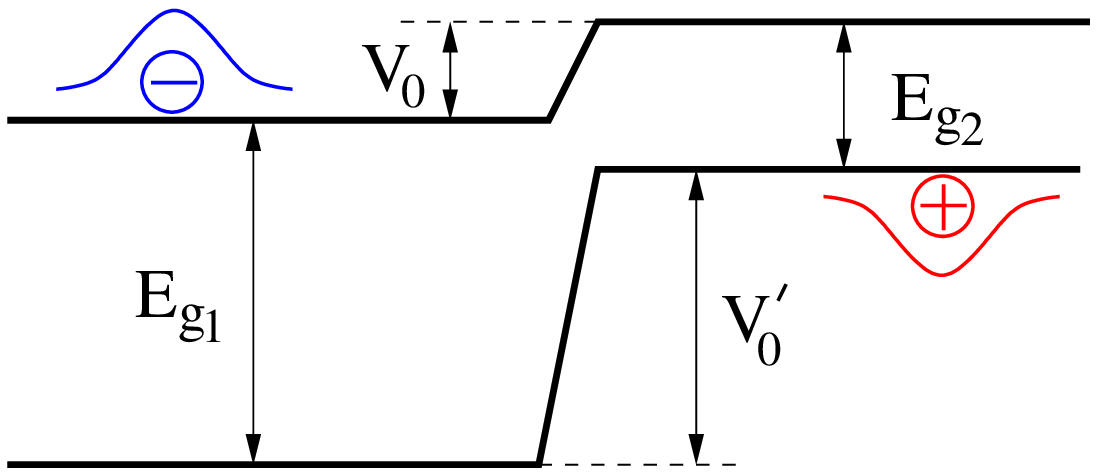} \\
\hspace*{0cm}(c)
\caption{(a) Zig-zag and (b) mesoporous type interfaces, inside the scattering region $\Omega_0$. The black (left hand side) regions correspond to 
the reference potential, while the yellow (right hand side) regions correspond to the band offsets.
(c) Band offsets for electrons ($V_0$) and holes ($V'_0$). 
}
\label{potentials}
\end{figure}

Using the R-matrix method the scattering wavefunctions can be calculated for an energy $E$,
which may vary continuously. Being solutions of the stationary Schr\"odinger equation (\ref{sch})
the functions $\Psi_\nu(\rr;E)$ form an orthogonal system:
\begin{equation}
\langle \Psi_\nu(\rr;E) | \Psi_{\nu'}(\rr;E') \rangle = \delta_{\nu\nu'} \delta(E-E').
\end{equation} 
In practical calculations, we discretize the energy axis and use a finite set of energies $\{E_k\}_k$. 

We assume the initial quantum state of a particle is given by $\Psi(\rr, t=0) \equiv \Psi_0(\rr)$,
typically, although not restricted to, a wavepacket inside the scattering region.
For a time independent scattering potential, the time evolution is described by:
\begin{equation}
\Psi(\rr,t) = \sum_{k,\nu} C_{k,\nu} \Psi_\nu(\rr;E_k) \exp\left(-i\frac{E_k}{\hbar}t\right),  
\end{equation}
where $C_{k,\nu} = \langle \Psi_\nu(\rr;E_k) | \Psi_0(\rr) \rangle$ are the expansion
coefficients of the initial wavefunction in the orthonormated basis made of the scattering functions.

\subsection{Computational method}

The main advantage of the R-matrix method relies on the fact that 
total computational cost is divided in two parts:
in the first part, the solution of the Wigner-Eisenbud problem $\{(\chi_l,E_l)\}_l$,
the transversal eigenvalue problem in each lead $\{\Phi_{\nu}(\rr^\perp_{s}),E_{\perp}^{\nu}\}_l$
and the calculation of the overlap integrals $(\chi_l)_\nu$ do not depend on the total energy $E$;
in the second part, the R-matrix, the S-matrix and optionally transmission functions and scattering
wavefunctions are determined for each total energy $E$.
Usually, for moderate number of channels and a large enough basis set, the Hamiltonian 
diagonalization performed in order to obtain the eigensystem $\{(\chi_l,E_l)\}_l$ represents 
the largest computational cost. This step is performed only once, while the energy dependent
calculations require a much smaller computational effort.  
The advantage of this method becomes particularly important when a relatively large number of transmissions and wavefunctions
need to be determined. 
\cblue{In addition, the transport properties may be interpreted using the Wigner-Eisenbud functions and energies \cite{manoles1}.}
In the following we detail the computational steps in our approach.

The rank of the Hamiltonian matrix in a three-dimensional model $(d=3)$ is $N_b = N_{bx} \times N_{by} \times N_{bz}$,
where $N_{bx}, N_{by}, N_{bz}$ are the numbers of basis functions corresponding to each spatial dimension.
The basis functions $\phi_{ijk}(\rr)=\phi_{xi}(x)\times\phi_{yj}(y)\times\phi_{zk}(z)$ defined in $\Omega_0$ should obey the vanishing normal derivative imposed by the condition (\ref{we2}) at the planar interfaces with the leads $\Omega_s$.
Although specific geometries require different basis sets, a generic model consists of identifying $\Omega_0$ with a parallelipipedic box of volume $L_x \times L_y \times L_z$. Assuming the origin of the coordinate system is in the middle of $\Omega_0$, we may choose
$\phi_{xi}(x)=1/\sqrt{L_x}\cos(i\pi(x+L_x/2)/L_x)$, with $i=1,2,\ldots$ and similarly for $\phi_{yj}(y)$ and $\phi_{zk}(z)$. By a one-time diagonalization, of typical cost ${\mathcal O}(N_b^3)$, the Wigner-Eisenbud functions and energies are determined. 
The transversal modes $\Phi_\nu(\rr^\perp_s)$ are found by solving Eq.\ (\ref{sch_perp}) in each lead $\Omega_s$, which are $(d-1)$-dimensional problems and therefore require a significantly lower computational cost, compared to the Wigner-Eisenbud problem. 

In the following one needs to set the maximum number of channels $n_s$ considered in each lead, the total number of channels $N_{\mbox{\scriptsize ch}} = \sum_s n_s$ being the rank of the R- and S-matrices, which are rigorously infinite matrices. The number of channels $n_s$ should be large enough to include at least the open channels and an appropriate number of closed ones, to ensure the convergence of the calculations, in particular of the transmission functions. 
At this point one can calculate explicitly the integrals $(\chi_l)_\nu$ from Eq.\ (\ref{chilnu}).
One should note that all these steps are energy independent.

In the second step, the R-matrix elements for a particular total energy $E$ are found by simple summations, of total cost 
${\mathcal O}(N_b \times N_{\mbox{\scriptsize ch}}^2)$. As the ${\rm k}$-matrix is diagonal, finding the
matrices [1$\pm\frac{i}{m^*}$Rk] from Eq.\ (\ref{smatrix}) is achieved by an insignificant cost of ${\mathcal O}(N_{\mbox{\scriptsize ch}})$.
The most demanding part is the computation of the inverse of the dense matrix [1$-\frac{i}{m^*}$Rk], which may be performed by standard LAPACK \cite{lapack} routines with scaling of ${\mathcal O}(N_{\mbox{\scriptsize ch}}^3)$. The subsequent matrix multiplication in Eq.\ (\ref{smatrix}) and the similarity transform used to obtain the $\tilde{\rm S}$ matrix do not exceed the computational cost of matrix inversion. These steps are performed for a number of $N_E$ total energies, so that the total cost is ${\mathcal O}(N_E \times N_{\mbox{\scriptsize ch}}^3)$. If $N_{\mbox{\scriptsize ch}} \ll N_b$, which is a typical situation, the computational burden mostly falls on solving the Wigner-Eisenbud problem. One should note that for an accurate representation of the
$\Phi_\nu(\rr^\perp_s)$ transversal modes, the number of channels considered in the calculations should be significantly smaller than the number of basis elements used for diagonalizing the leads Hamiltonian given by 
Eq.\ (\ref{sch_perp}). 
%in $\Omega_0$ on the corresponding spatial directions.  
The total cost of finding the transmission functions is 
${\mathcal O}(N_b^3 + N_E \times N_{\mbox{\scriptsize ch}}^3)$
and one can easily see that the cost of energy dependent part becomes comparable to the Wigner-Eisenbud problem only if the transmissions are computed for a large number of total energies, making the procedure effective in providing rapid varying transmission functions.

%The wavefunctions are calculated separately for each lead and the scattering region, satisfying the continuity relation at each interface $\Gamma_s$.

One should point out that the R-matrix method provides the set of scattering functions 
$\{\Psi_\nu(\rr;E_k)\}_{\nu,k}$ of the open quantum system, 
which are continuous over the entire interval, leads and scattering region.
One advantage of our method is the possibility of including potential
barriers at the lead/scattering region interfaces, of arbitrary height, and
therefore to tune the contacts to the leads from fully transparent
for electrons (or reflectionless) with no barrier,  to almost
opaque with high barriers.  An even bigger advantage is the fully
transparent case itself, which practically describes an infinite system
with a finite Hamiltonian matrix in the scattering region and asymptotic
wave functions outside.  Alternatively, a large system can in principle
be described as a finite, closed system, in a basis of wave functions
vanishing at the boundaries. But for practical calculations the size of
the basis must increase with the size of the system, eventually becoming
prohibitive. Therefore, for given physical parameters and simulation time
our method is more efficient.  While the solution of the Wigner-Eisenbud
problem represents usually the most demanding part, the presence of the
leads generates only a minimal additional computational effort.
\cblue{Other methods are known in the transport theory, such as 
the Lippmann-Schwinger equation \cite{lippmann}, 
the quantum boundary transmitting method (QBTM) \cite{lent}
and the non-equilibrium Green's functions (NEGF) formalism \cite{kadanoff}.
The Lippmann-Schwinger equation written in coordinate representation 
is an integral equation of Fredholm type, which can be solved numerically 
by discretization at a given total energy. QBTM is another wavefunction-based method, which gives a real 
space solution of the Schr\"odinger equation using finite difference or finite elements 
approach. In this case solving a linear system of equations which depends on the total 
energy is required.
} \cblue{ While the LS and QBTM methods describe ballistic transport, the NEGF formalism
 is a comprehensive tool which may include inelastic scattering as well and involves the 
calculation of the retarded/advanced Green's functions for a given total energy. 
In contrast to the R-matrix method, where the diagonalization of the Hamiltonian in the scattering region
is performed only once,
the aforementioned methods require a separate calculation, which must be initiated for each total energy. 
In this context, the R-matrix method becomes overall more efficient in computing the transmission functions or the scattering wave functions for a large energy set. 
}

\section{Application to nanostructured interfaces}

\subsection{Physical systems}

Nanostructured inhomogeneous interfaces increase the light scattering and the device active area in photovoltaic 
applications, enhancing the generation of 
photoexcited carriers. The size of the inhomogeneities should be comparable with the light wavelength, 
i.e. hundreds of nanometers. Different techniques have been used to achieve this goal, e.g. by using nanowire
arrays or mesoporous materials. The latter have been increasingly used for boosting the solar cell 
efficiencies. As concrete examples, one can mention thin films of 
nearly spherical TiO$_2$ or ZnO aggregates with typical sizes up to a micron.
Furthermore, the aggregates possess inner structure, being composed of nanocrystallites, 
with typical sizes of $\sim$15nm \cite{zhang}. 
Given the quite different length scales involved in the physical structures of interest and the 
operation conditions, hybrid (quantum-classical) transport models should be used:
at a few nanometer length scales, e.g. at the nanocrystallite scale, 
a quantum mechanical transport model should be best suited, while at the
aggregate level and larger scales a drift-diffusion model should be employed. 

\begin{figure}[t]
\centering
\includegraphics[width=4.cm]{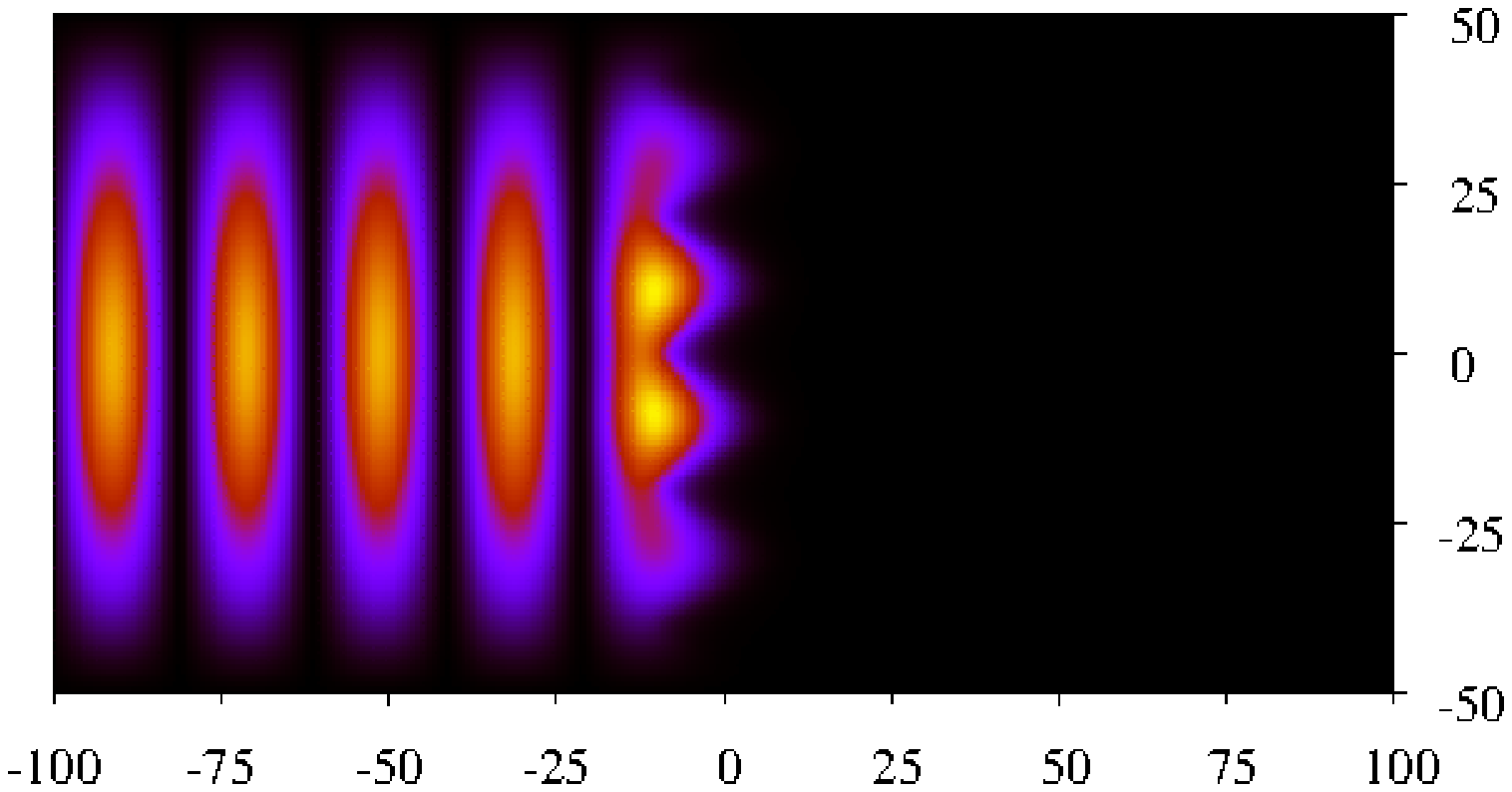}
\includegraphics[width=4.cm]{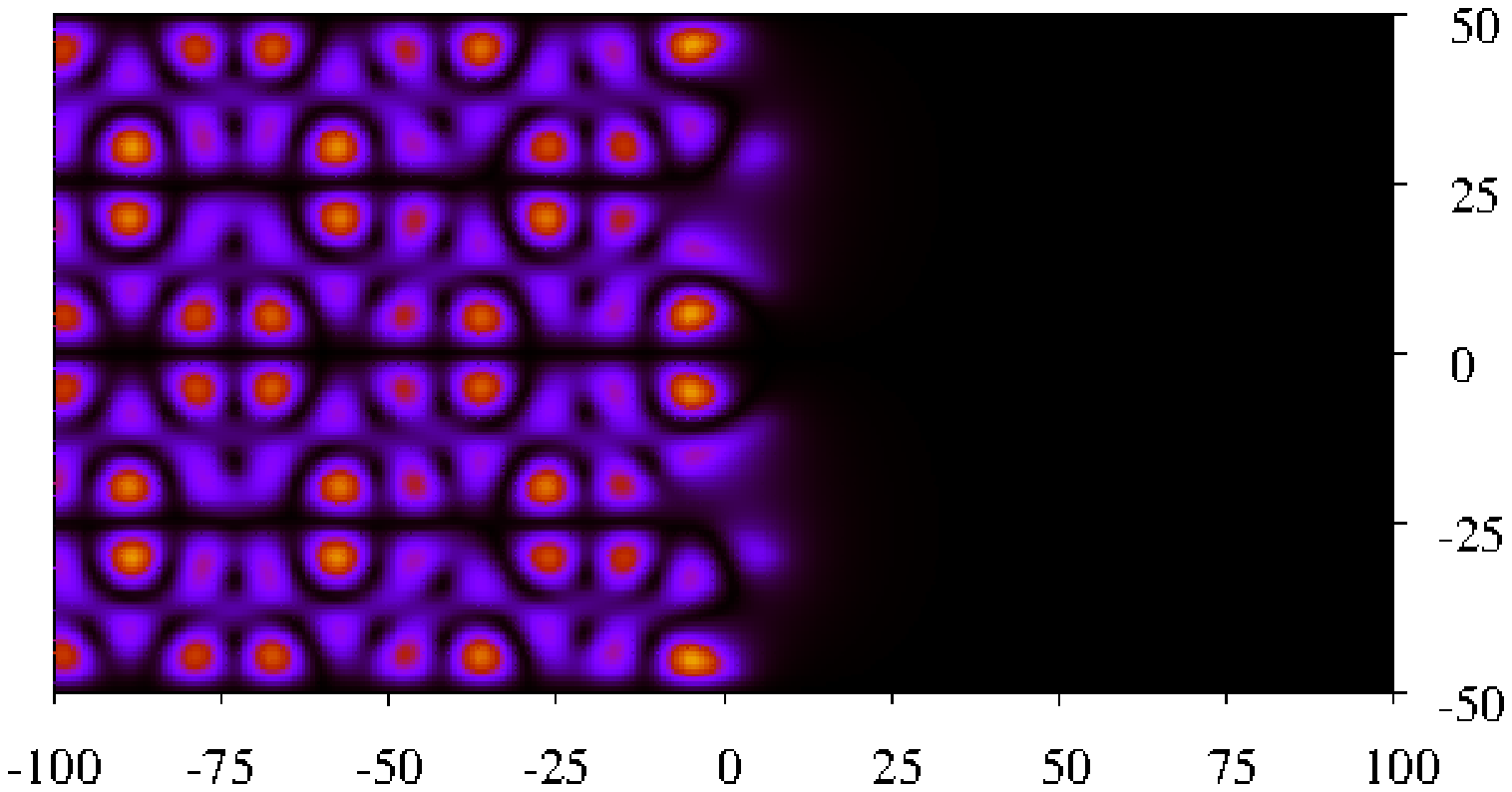} \\
\includegraphics[width=4.cm]{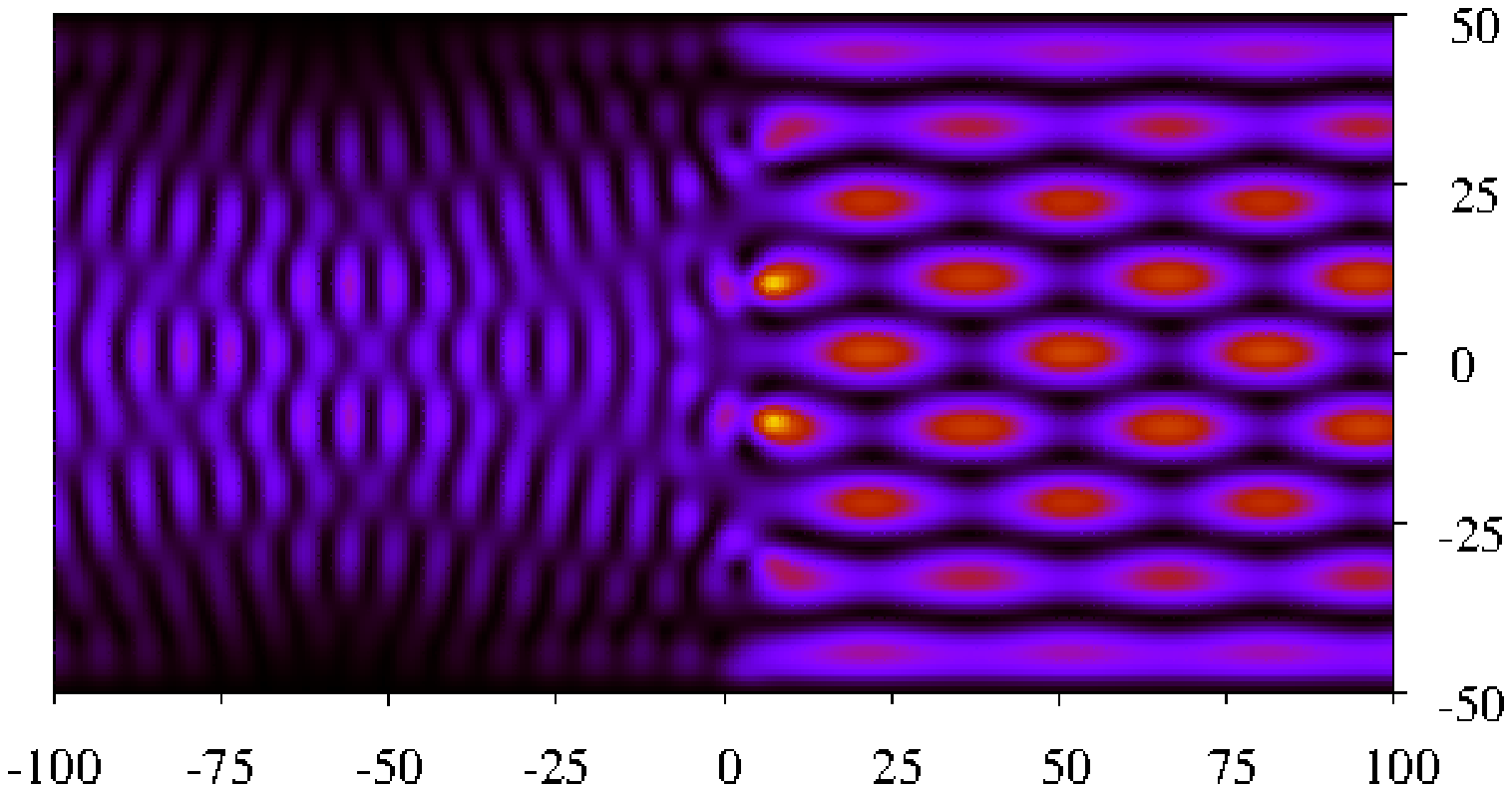} 
\includegraphics[width=4.cm]{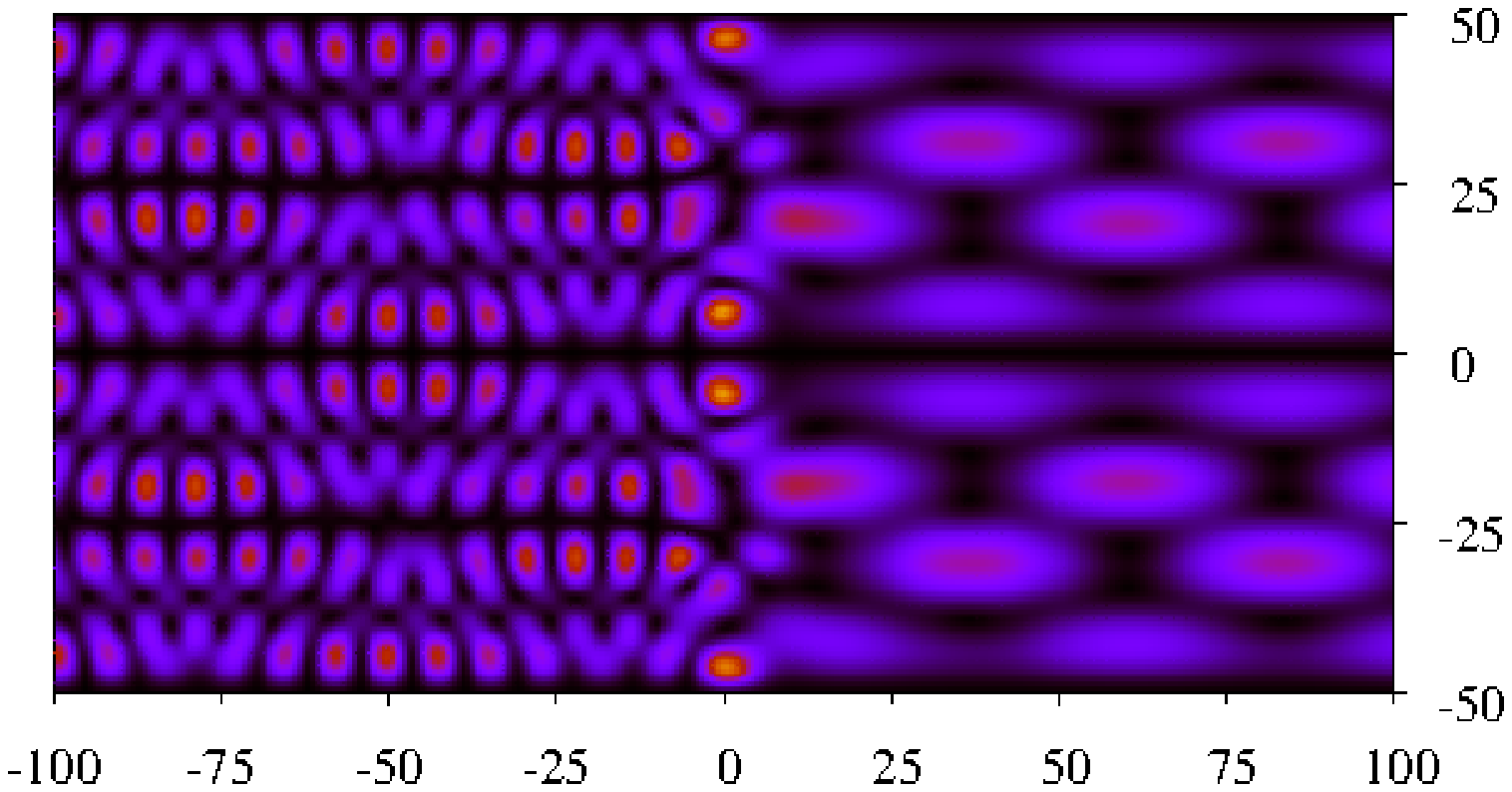} \\
\hspace*{0cm}(a)\hspace*{3.7cm}(b)
\caption{Zig-zag interface: scattering wavefunctions (absolute value square) for
 (a) lowest transverval mode for total energies $E = 0.015$ eV (upper), $0.15$ eV (lower) and 
(b) the 8th mode for $E = 0.09, 0.15$ eV. The bandoffset potential is $V_0 = 0.1$ eV.
The images include the scattering region and portions of the leads as described in the text.
The scattering region corresponds to the interval $(-50,50)$ nm on the horizontal direction. 
}
\label{wfs_zigzag}
\end{figure}

\begin{figure}[t]
\centering
\includegraphics[width=4.cm]{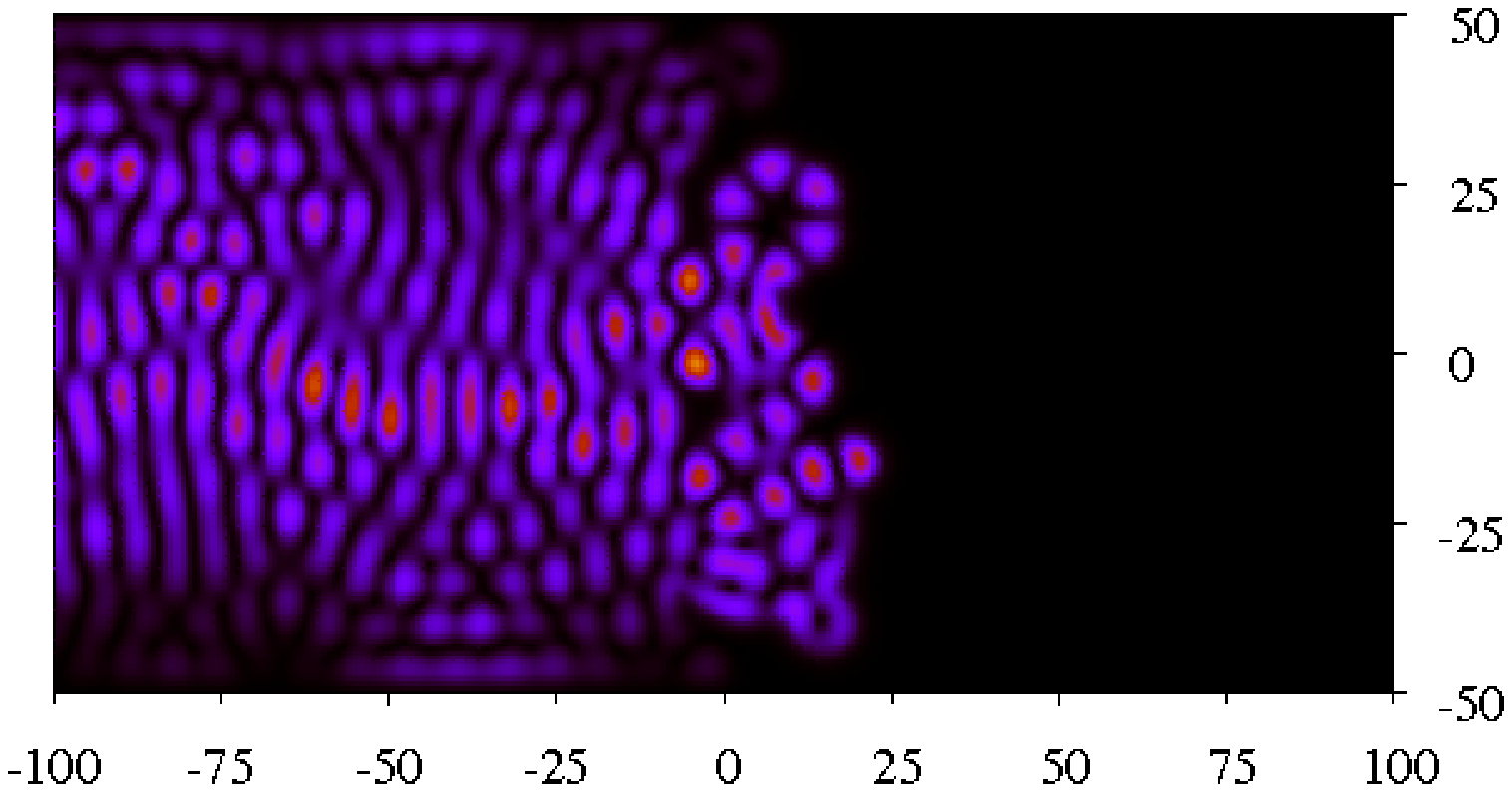}
\includegraphics[width=4.cm]{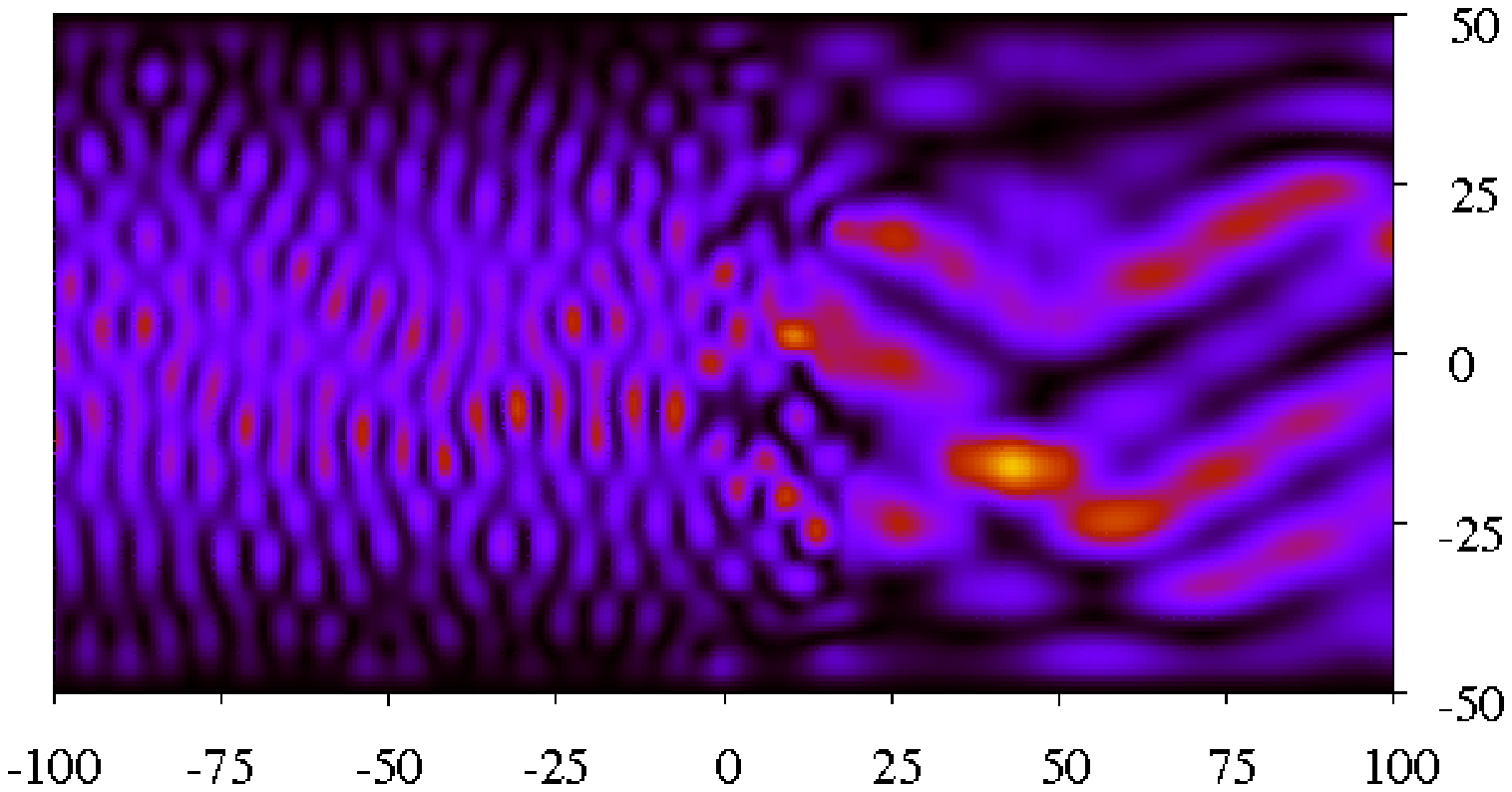} \\
\includegraphics[width=4.cm]{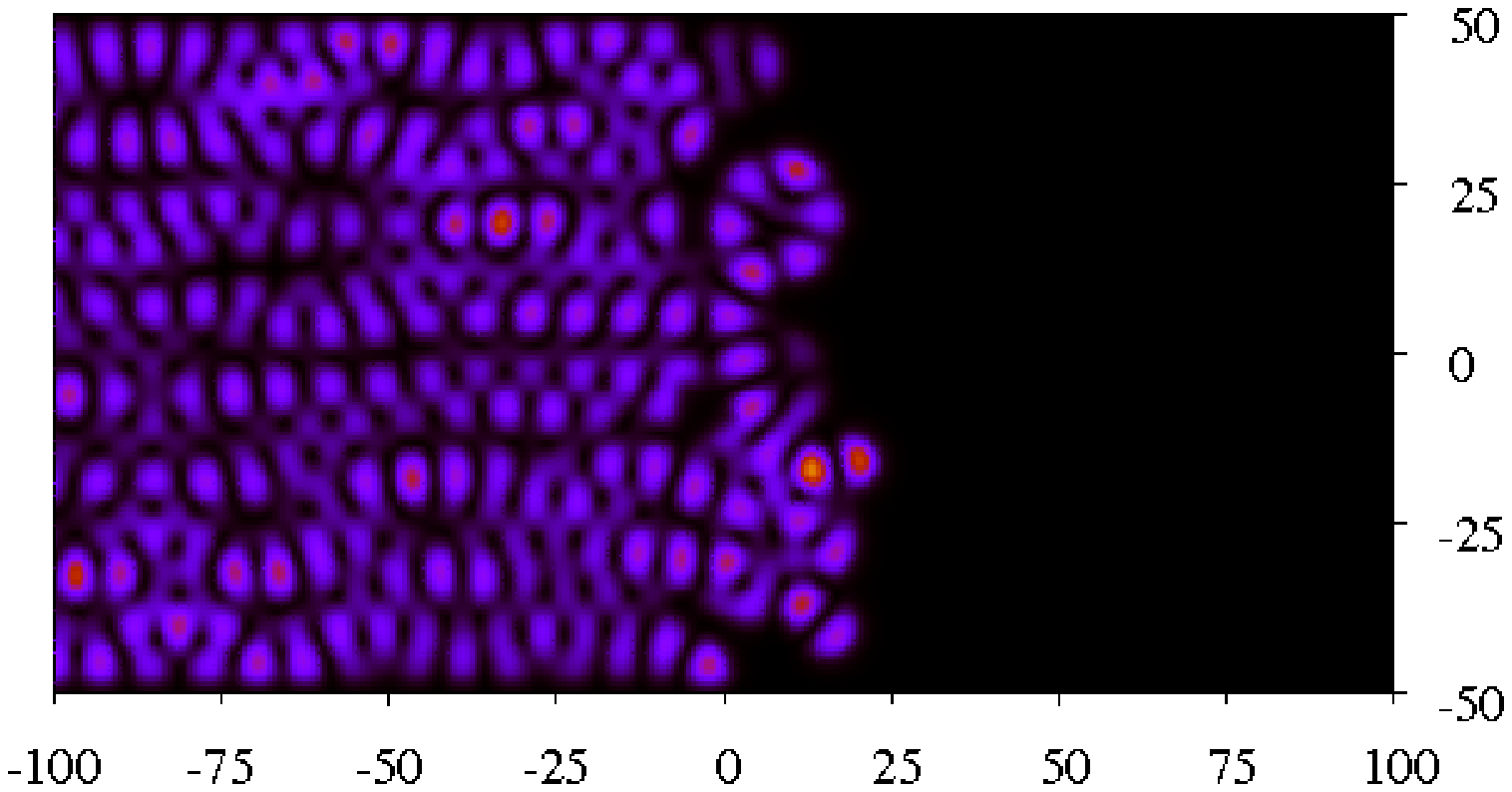}
\includegraphics[width=4.cm]{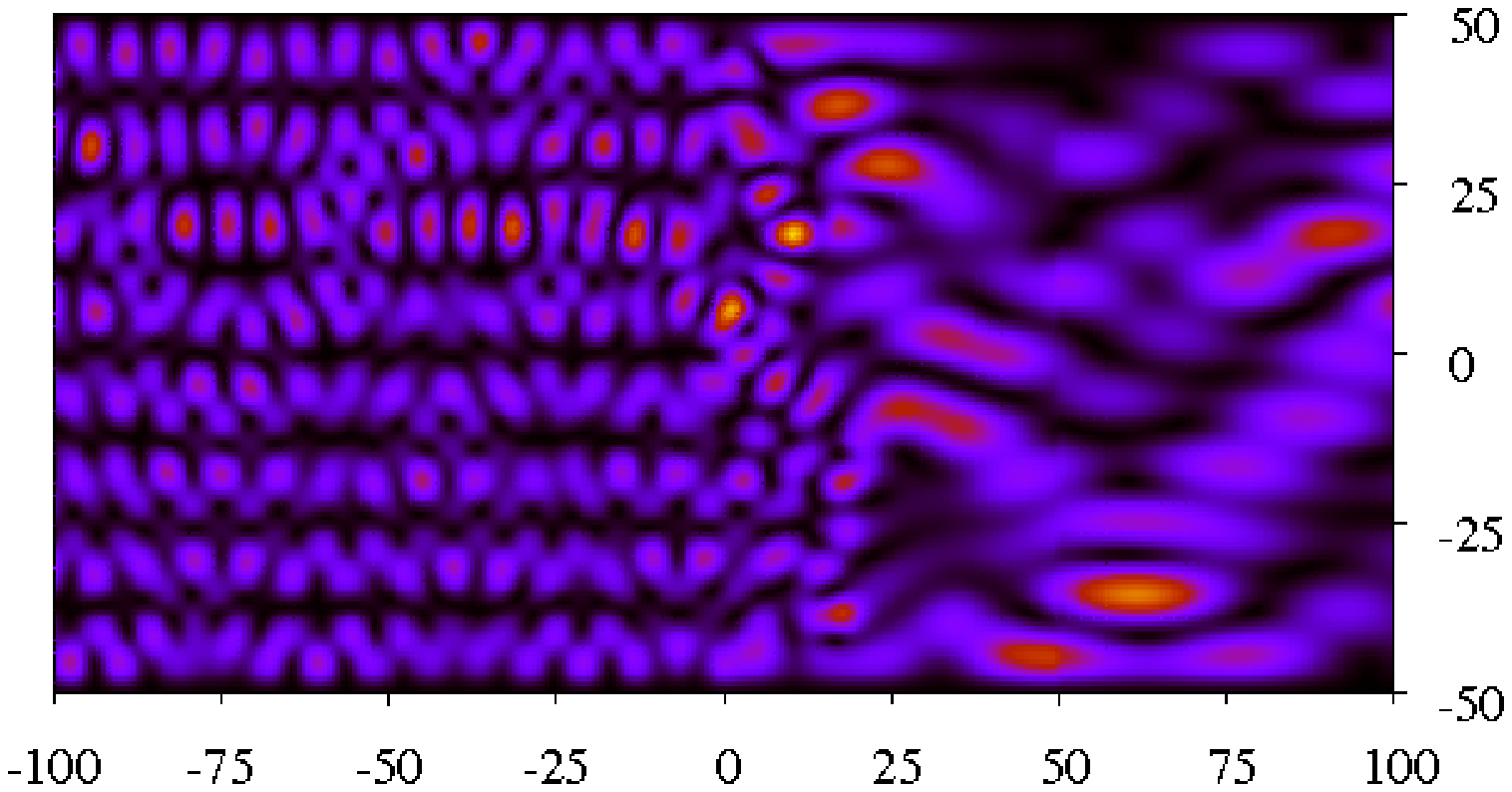} \\
\hspace*{0cm}(a)\hspace*{3.7cm}(b)
\caption{Mesoporous interface: scattering wavefunctions (absolute value square) for 
(a) 1st and 8th modes with $E = 0.18$ eV (upper) and $V_0 = 0.5$ eV (lower); (b)
the same modes and total energy, but for a lower bandoffset, $V_0 = 0.1$ eV. 
}
\label{wfs_mesoporous}
\end{figure}

We consider here two different types of nanostructured interfaces, namely the 
{\it zig-zag} and the {\it mesoporous} interface, as indicated in Fig.\ \ref{potentials}.
The first interface presents sharp regular variations, while the second one mimics a prototypical 
mesoporous layer.  
Flat-band condition is assumed for each of the two materials in the bulk, while at the interface 
there is a band offset $V_0$. 
Both types of structures introduce a local confinement potential for electrons and holes.
From the holes perspective, the potentials in Fig.\ \ref{potentials} are simply interchanged,
provided the bandgap is the same in the two materials. Generally, for different bandgaps in the two
materials, $E_{g1}$ and $E_{g2}$, the band offset for holes is $V'_0 = E_{g1}-E_{g2}+V_0$.  
Therefore, in a non-interacting one-particle picture, we may describe within the same formalism
the independent propagation of electrons and holes across the nanostructured interface \cite{nemnes7}.

\subsection{Scattering wavefunctions}

\begin{figure}[t]
\centering
\includegraphics[width=4.cm]{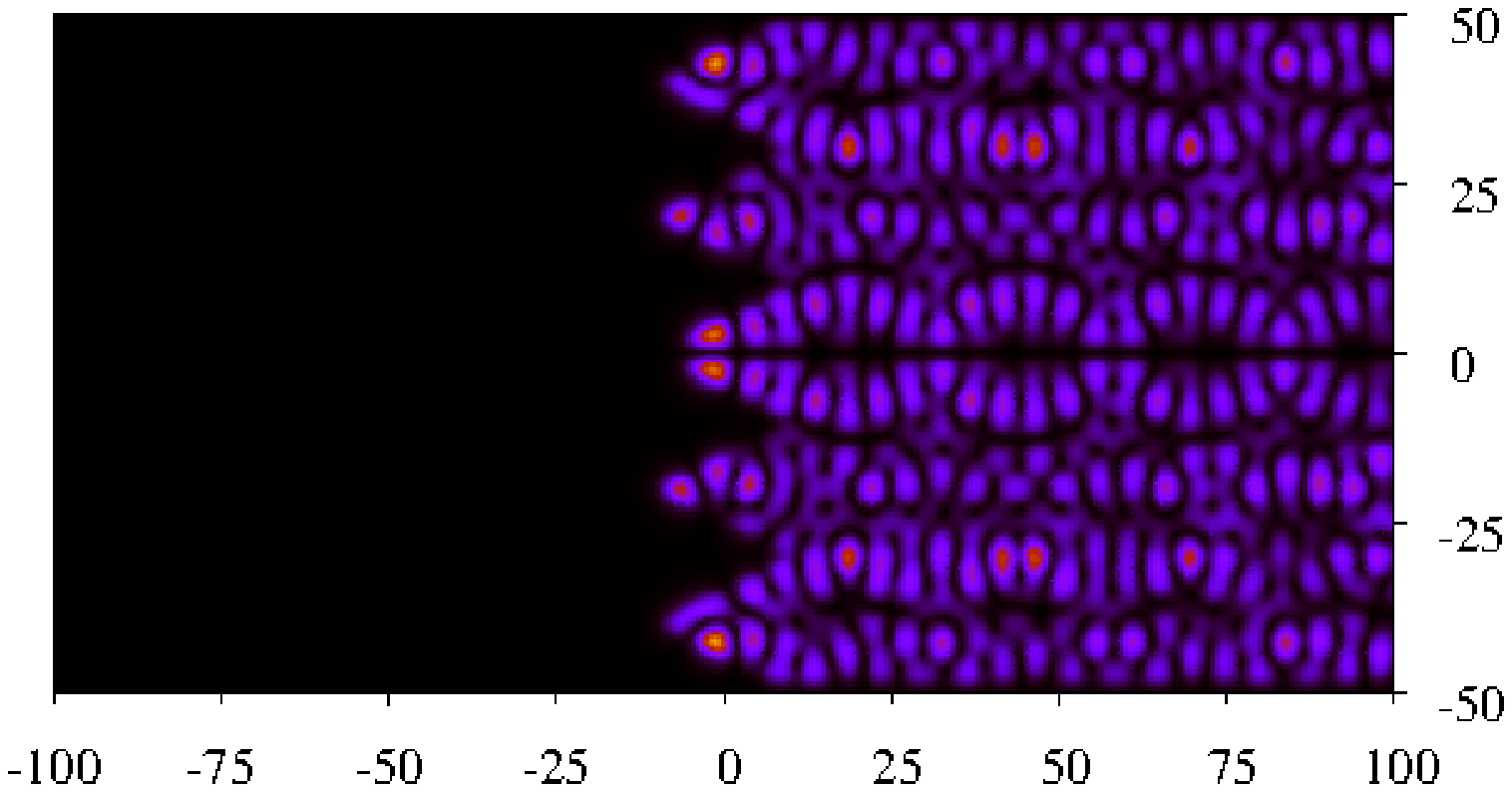}
\includegraphics[width=4.cm]{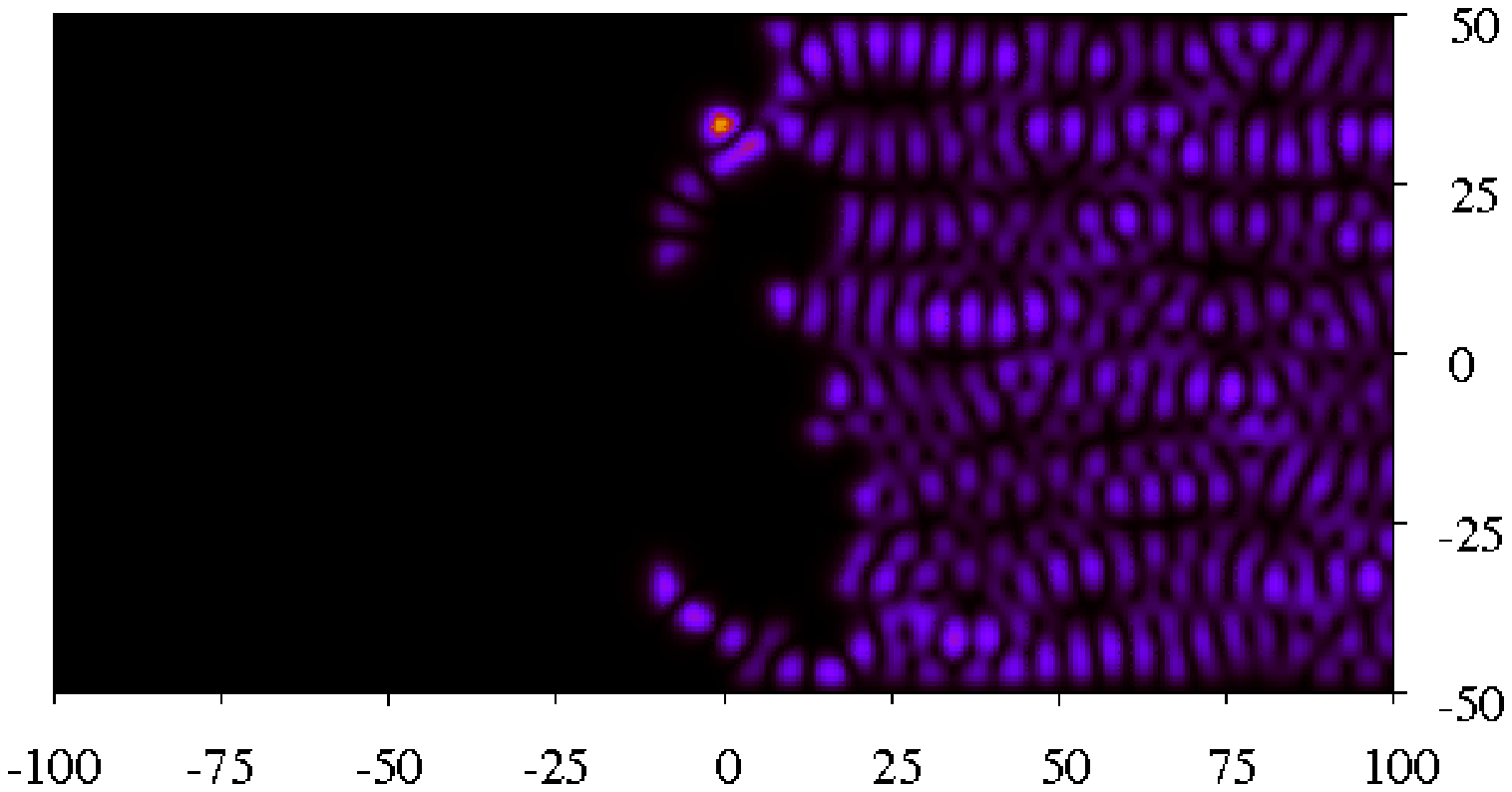} \\
\hspace*{0cm}(a)\hspace*{3.7cm}(b)
\caption{Scattering functions (absolute value square) corresponding to holes incoming from right, 
for the zig-zag (a) and mesoporous (b) interfaces.
The 8th mode is represented at $E=0.3$ eV, for a band offset $V'_0 = 0.5$ eV. 
}
\label{wfs_holes}
\end{figure}

We first analyze the solution of the stationary Schr\"odinger equation (\ref{sch}), subject to scattering boundary 
conditions for the two types of interfaces.  

Typical scattering wavefunctions for electrons are depicted in Figs. \ref{wfs_zigzag} and \ref{wfs_mesoporous} 
for the zig-zag and the mesoporous interface, for
two different transversal modes, 1st and 8th, and different total energies and band offsets. 
The wavefunctions are shown for the combined system of leads ($\Omega_s$) and scattering region ($\Omega_0$).
We consider a rectangular scattering region defined by the lengths $L_x=L_y=100$ nm 
as shown in Fig.\ \ref{potentials}
and additional portions of the leads of length $L_s=50$ nm are also depicted. 
At energies lower than the band offsets, there is total reflection, since the $V_0$ potential step extends in 
the right lead to infinity. 
At higher energies, transmitted wavefunctions are observed in the right lead, together with partial reflections
into the left lead.
As one can see from Fig.\ \ref{wfs_zigzag} the regular shaped zig-zag interface produces uniform patterns
 in the scattering functions and therefore in the charge localization.
On the other hand, in the case of a disordered mesoporous interface, as one may expect, asymmetries due to randomness are  
present. The shapes of the scattering potential can be correlated to the maxima of 
localization, which indicates the feasibility of our approach in the context of describing 
fine-grain nanostructured interfaces.
We also refer the reader to Ref.\ \cite{nemnes7} where sharp dendritic interfaces were analyzed.  

In the effective mass model, the hole perspective is mirrored with respect to the electron picture. 
Figure \ref{wfs_holes} shows the scattering wavefunctions for holes, which are incoming from the right hand side.
For the chosen parameters, $E=0.3$ eV and $V'_0=0.5$ eV, the holes are completely reflected and the two 
localization probability maps become complementary to the ones obtained for electrons: for the zig-zag interface, 
see Fig.\ \ref{wfs_zigzag}(b) (upper part) and for the mesoporous interface, see Fig.\ \ref{wfs_mesoporous}(a) 
(lower part).

\subsection{Wavepacket propagation}

\begin{figure}[t]
\centering
\includegraphics[width=2.0cm]{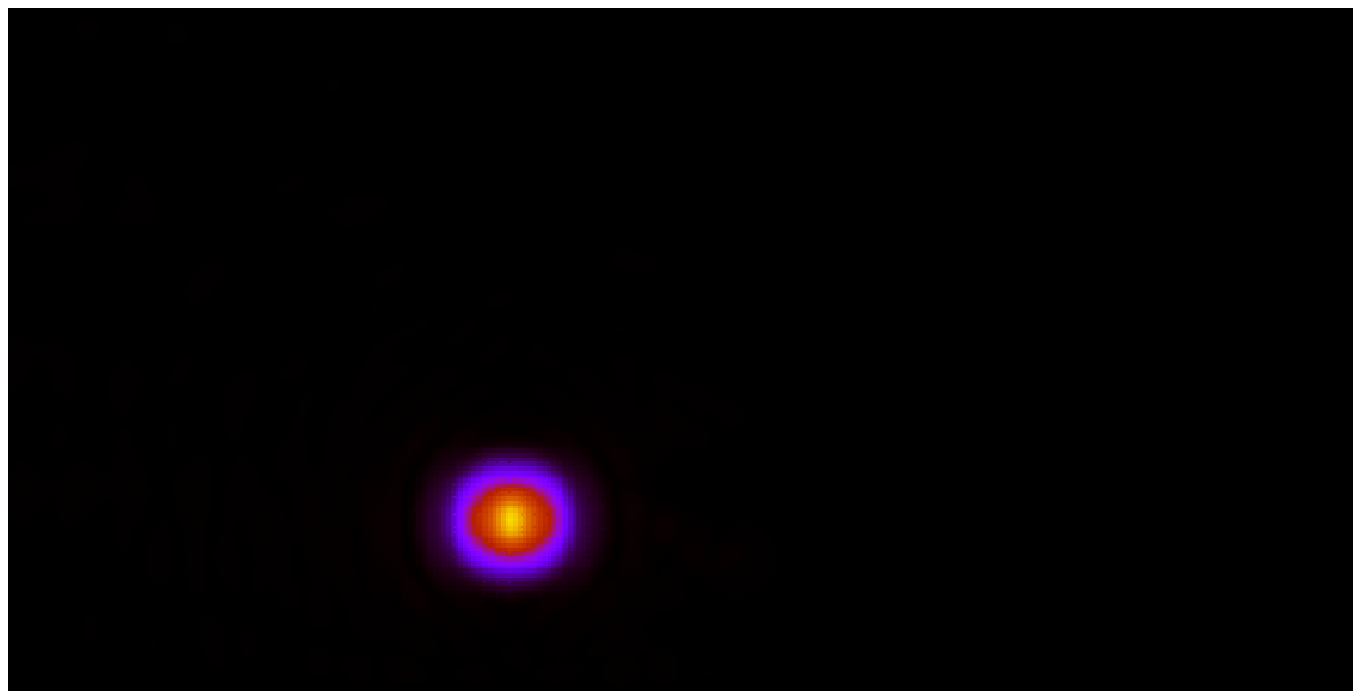}
\includegraphics[width=2.0cm]{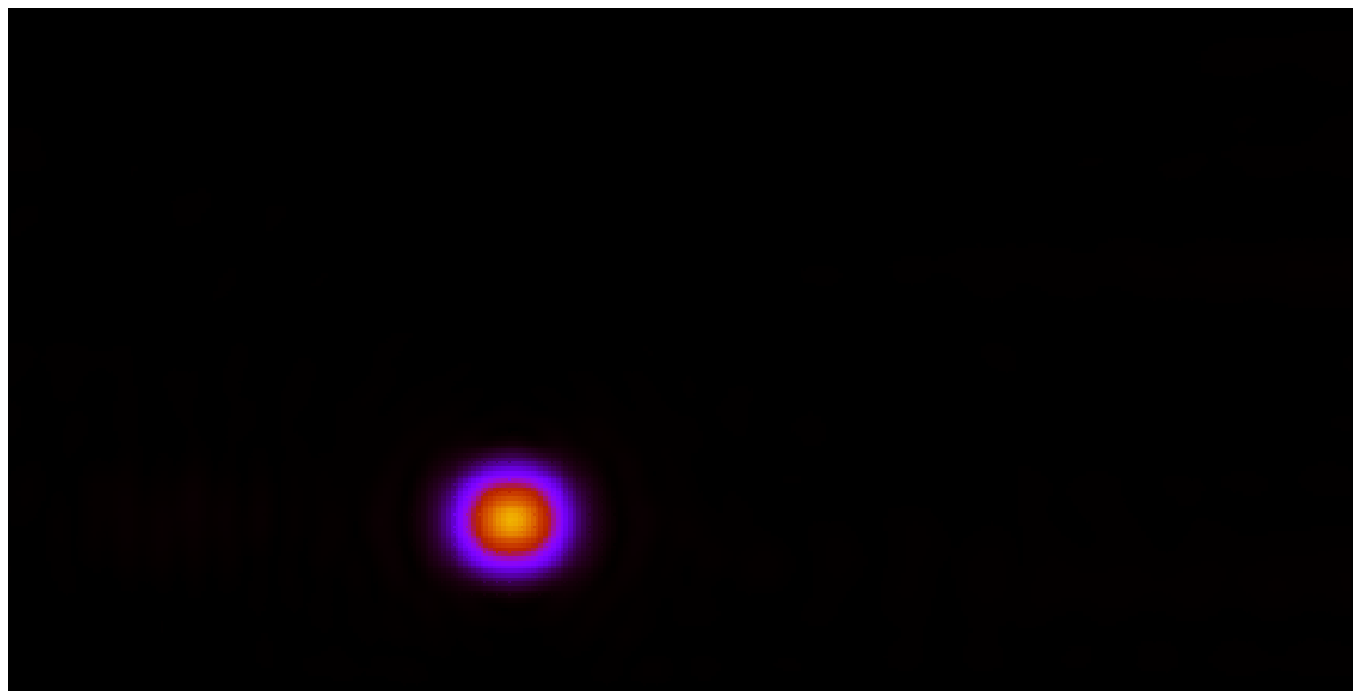} 
\includegraphics[width=2.0cm]{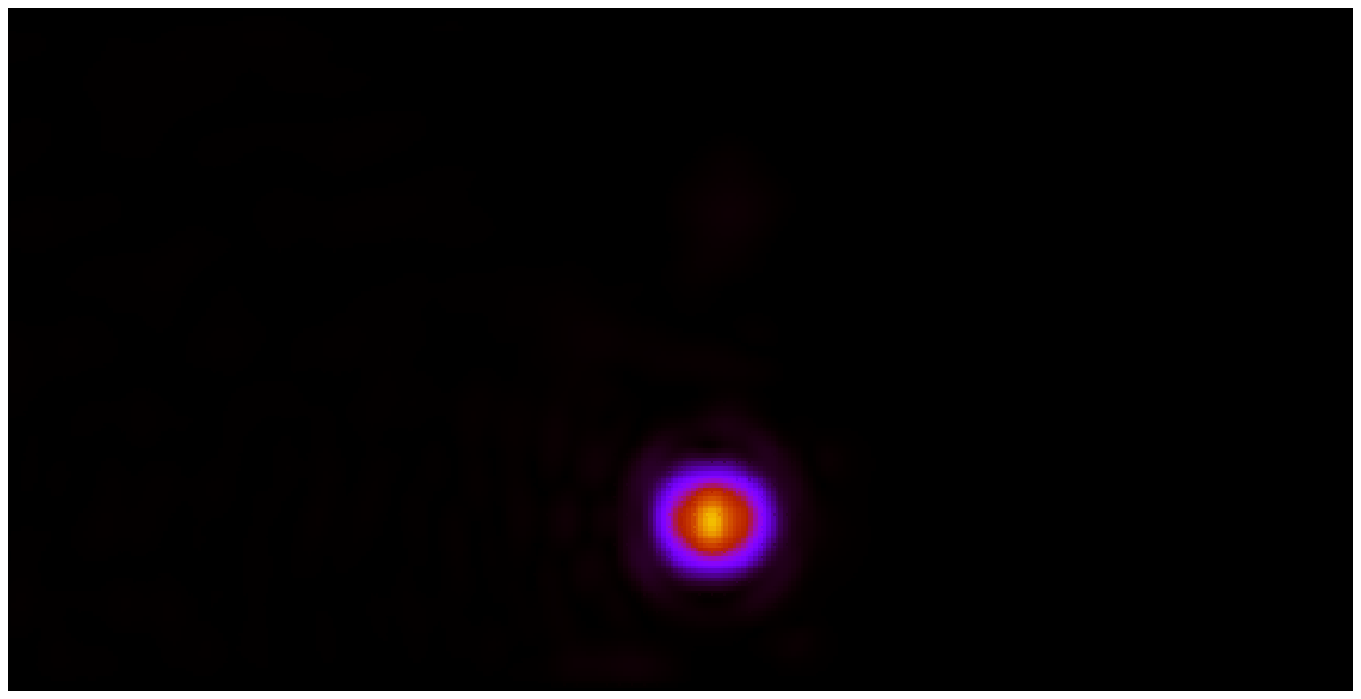}
\includegraphics[width=2.0cm]{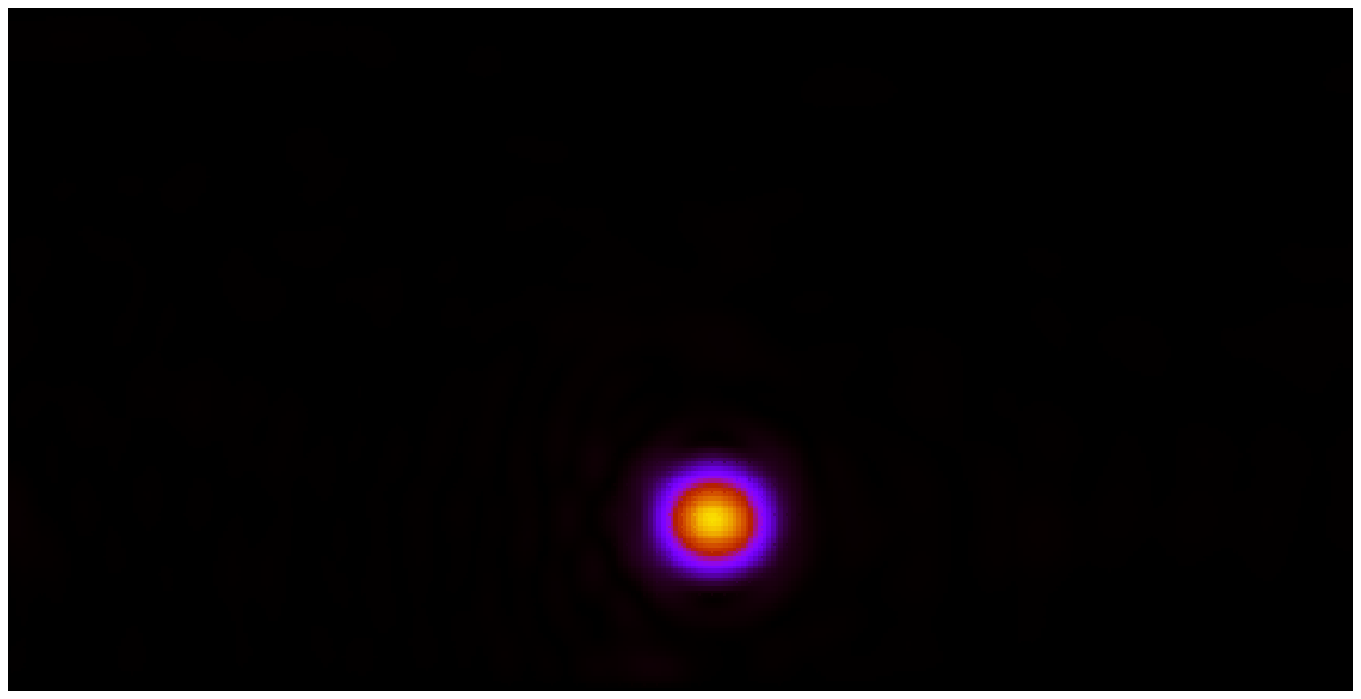} \\
\includegraphics[width=2.0cm]{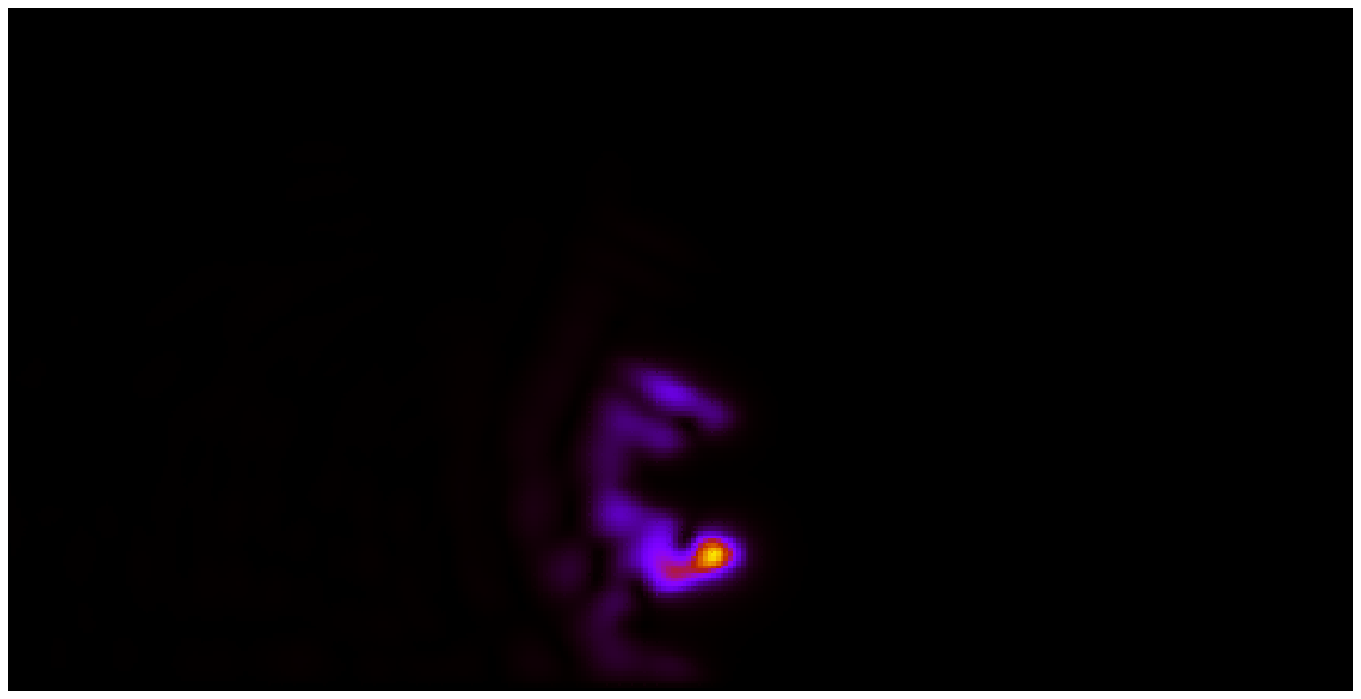}
\includegraphics[width=2.0cm]{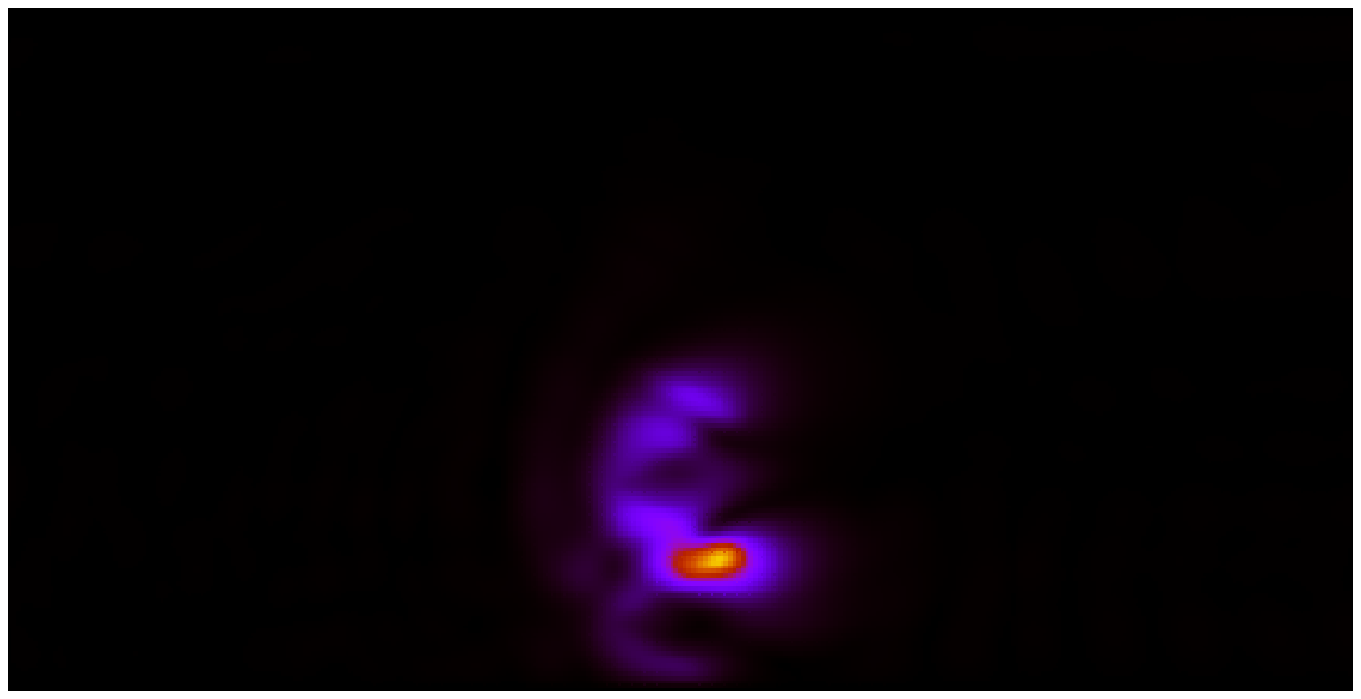} 
\includegraphics[width=2.0cm]{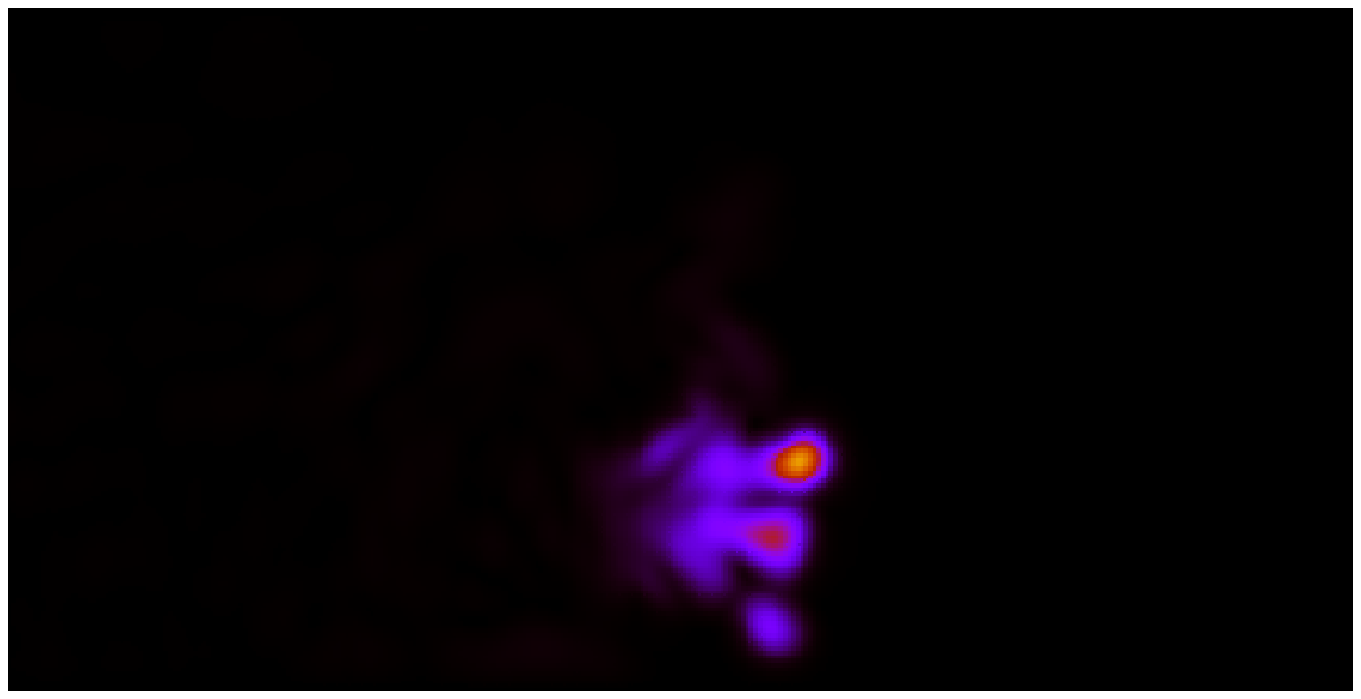}
\includegraphics[width=2.0cm]{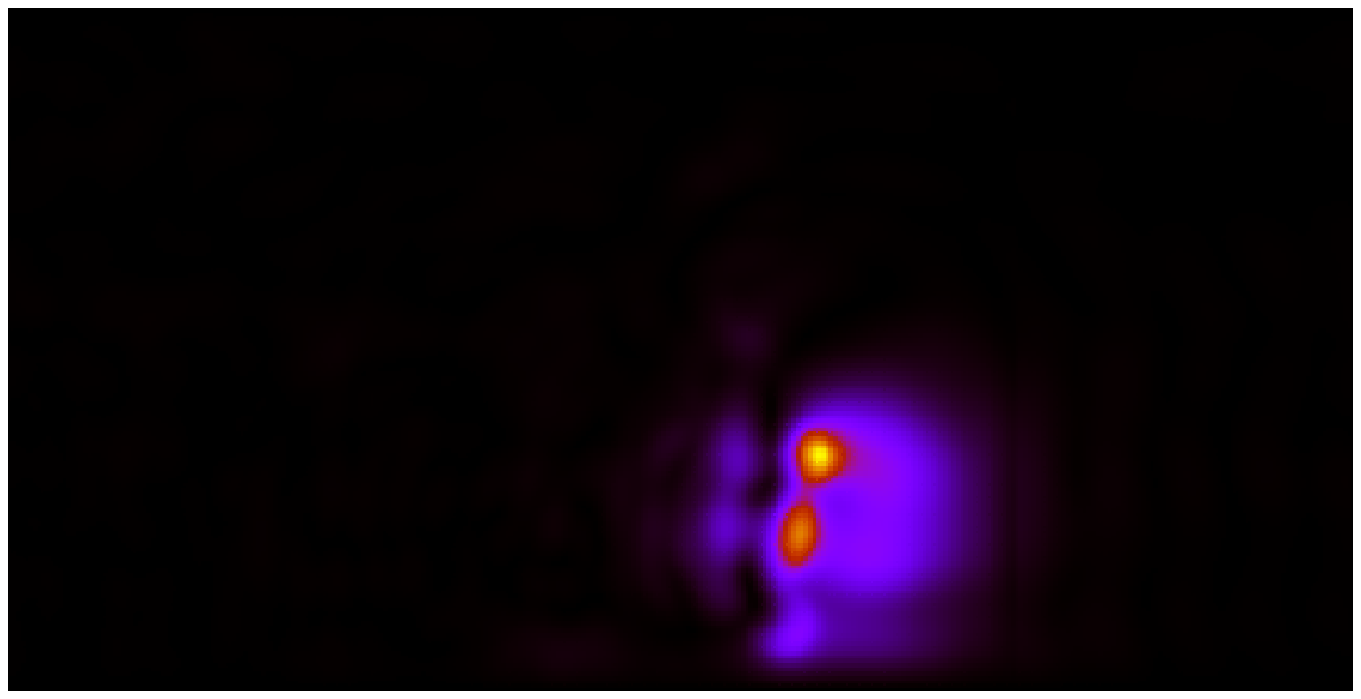} \\
\includegraphics[width=2.0cm]{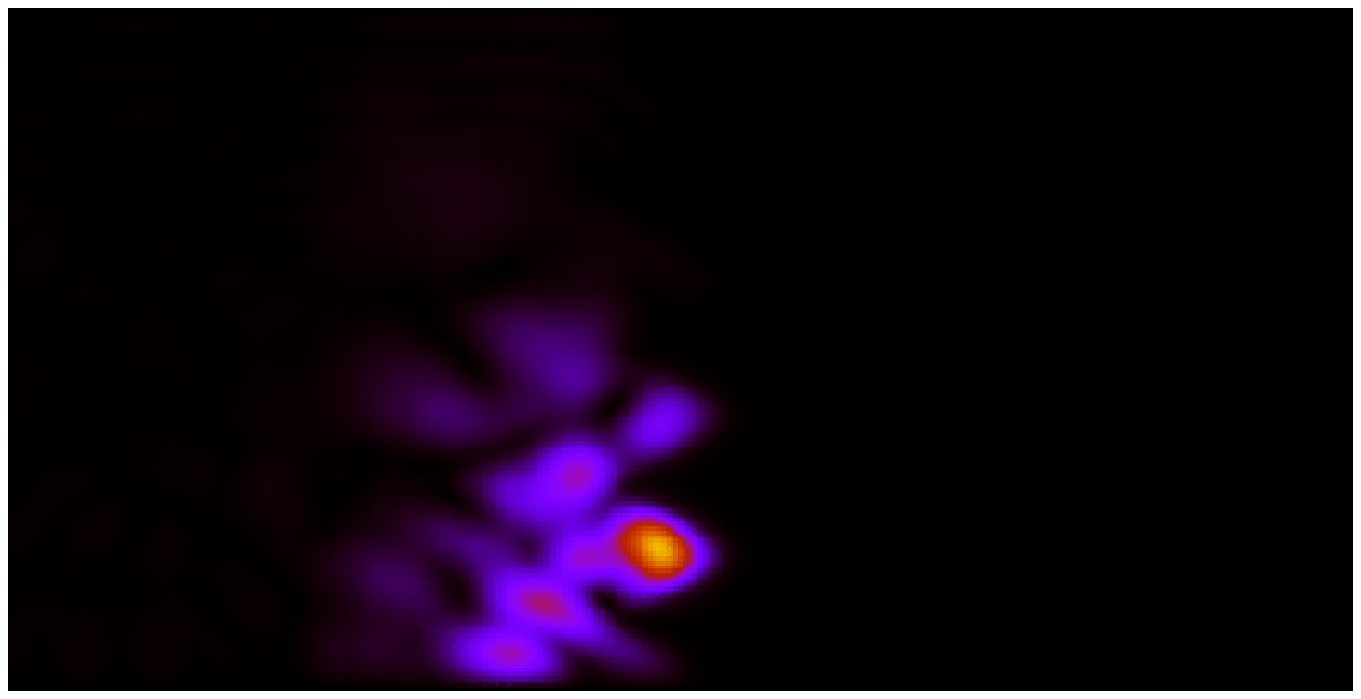}
\includegraphics[width=2.0cm]{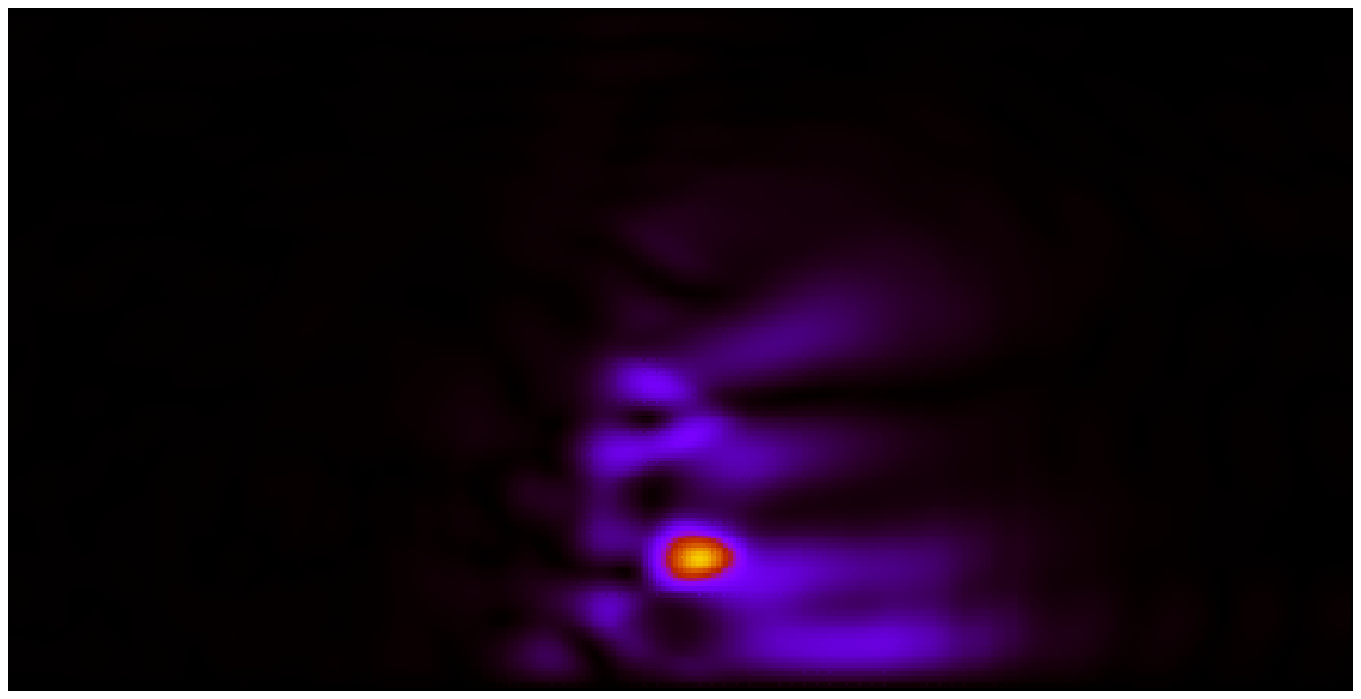} 
\includegraphics[width=2.0cm]{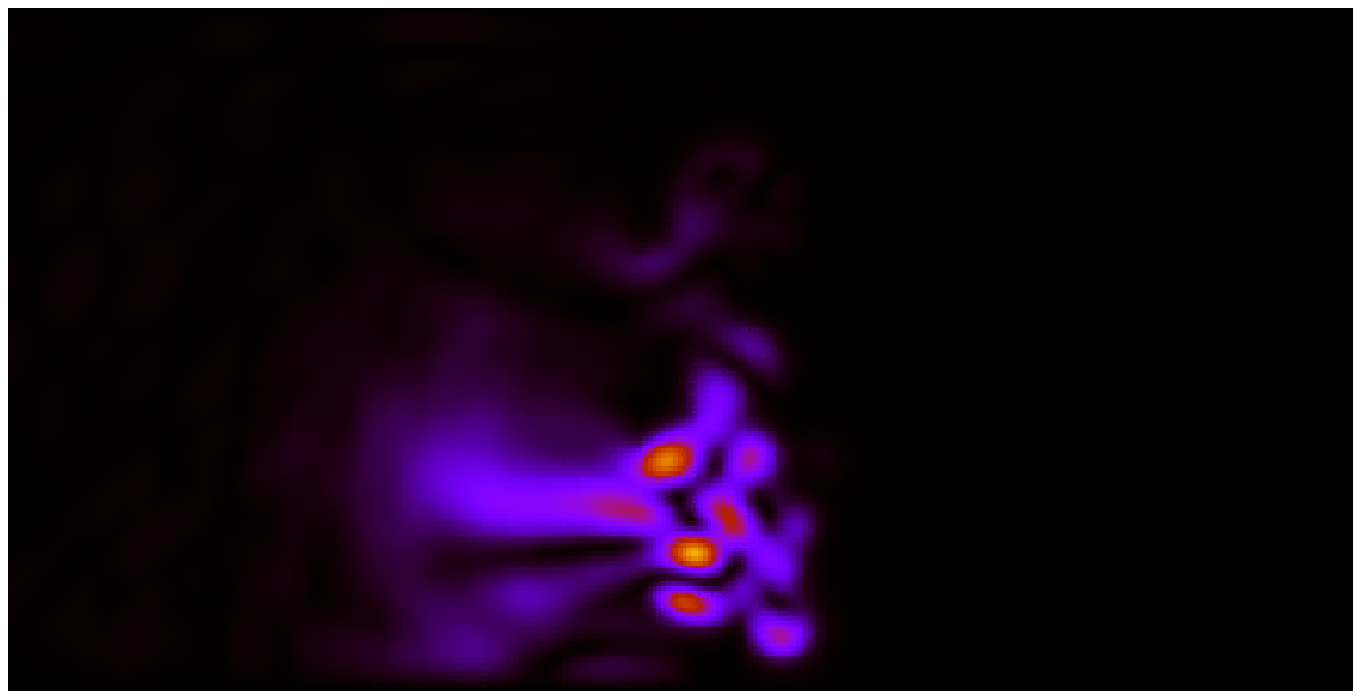}
\includegraphics[width=2.0cm]{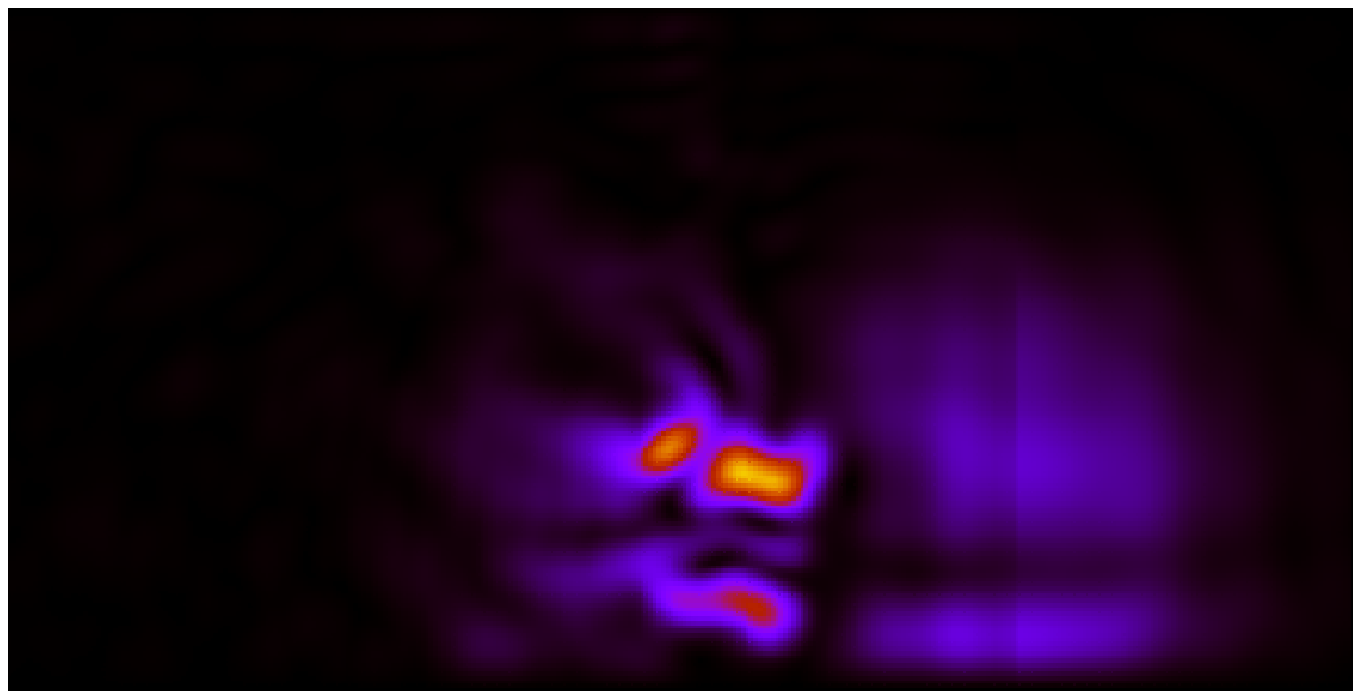} \\
\includegraphics[width=2.0cm]{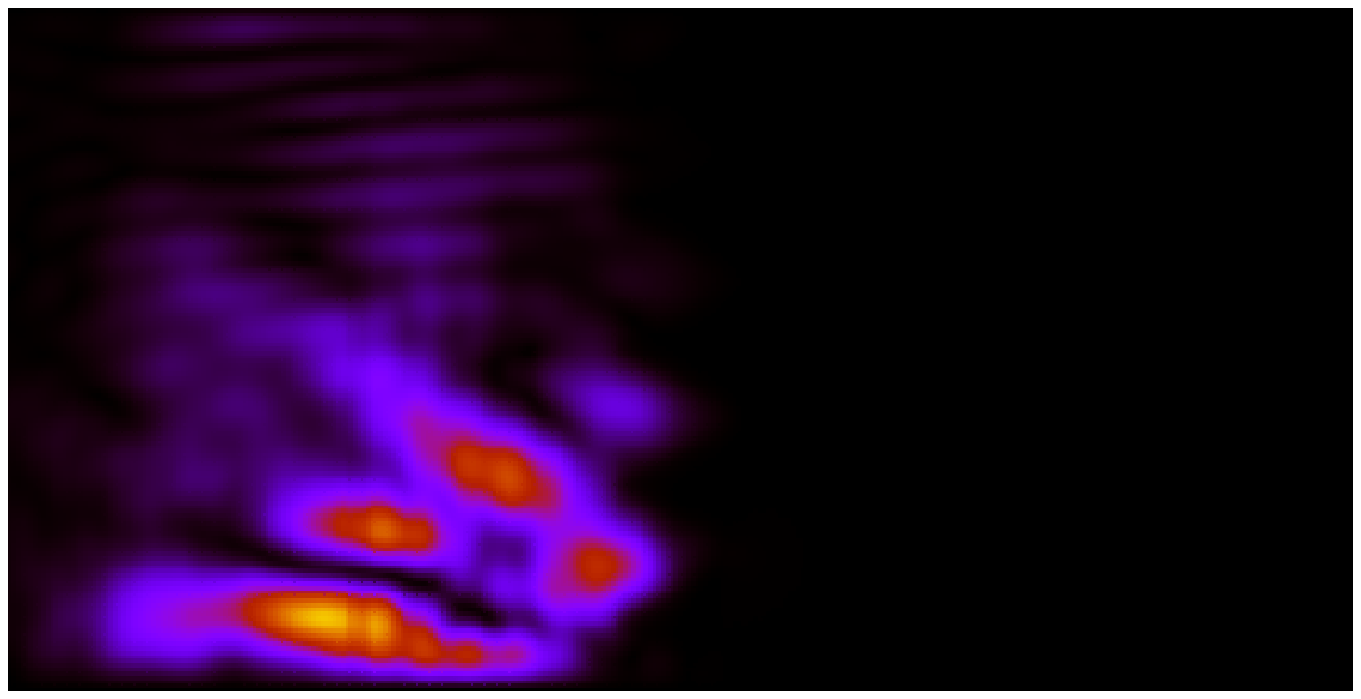}
\includegraphics[width=2.0cm]{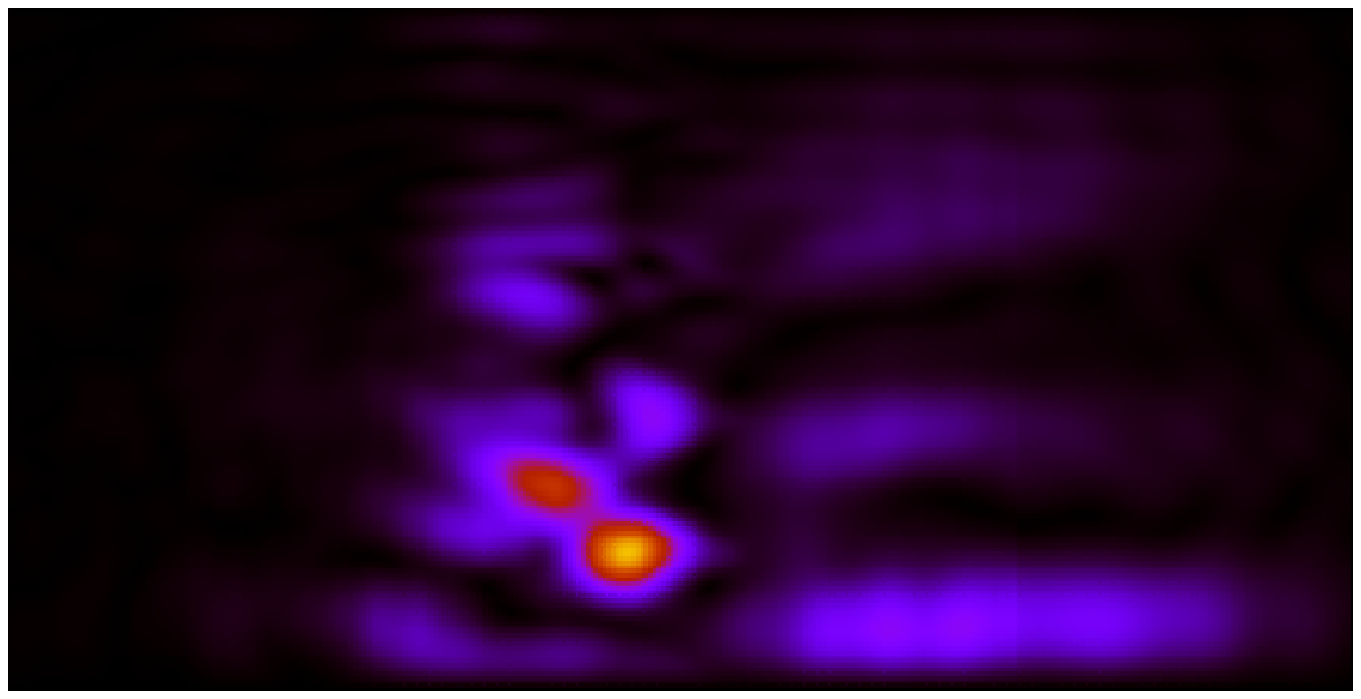} 
\includegraphics[width=2.0cm]{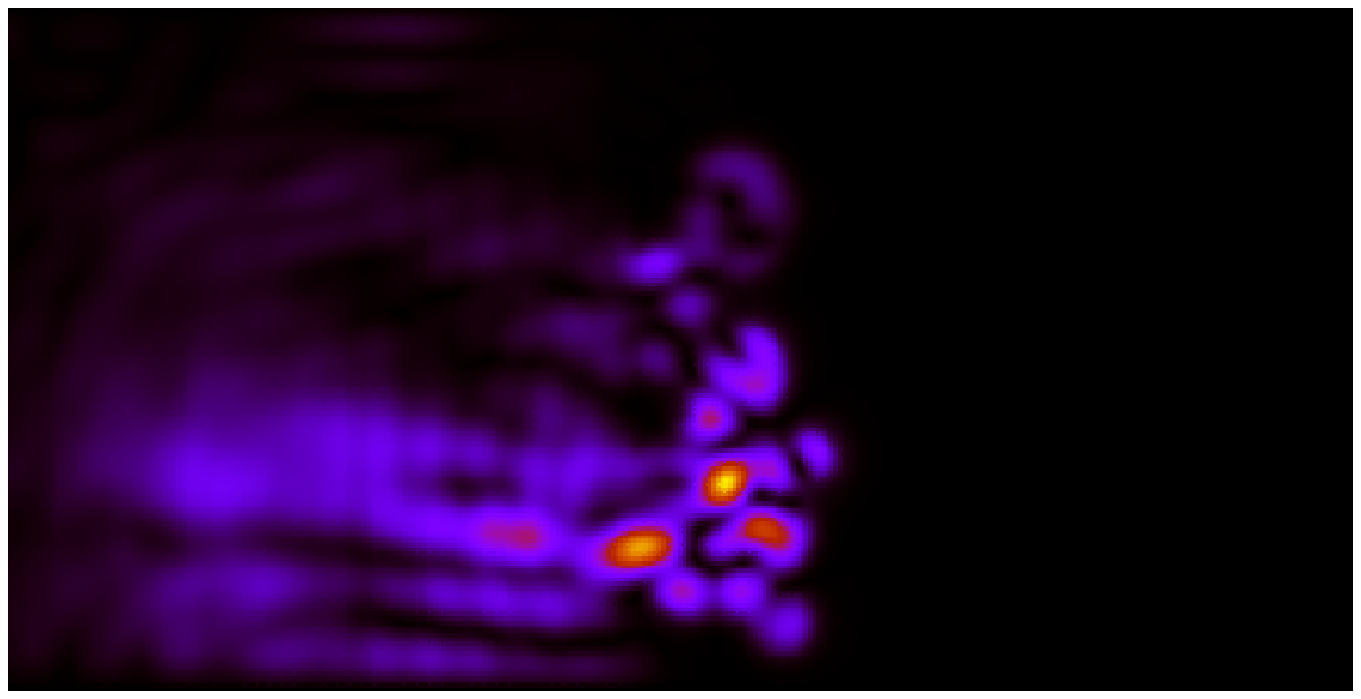}
\includegraphics[width=2.0cm]{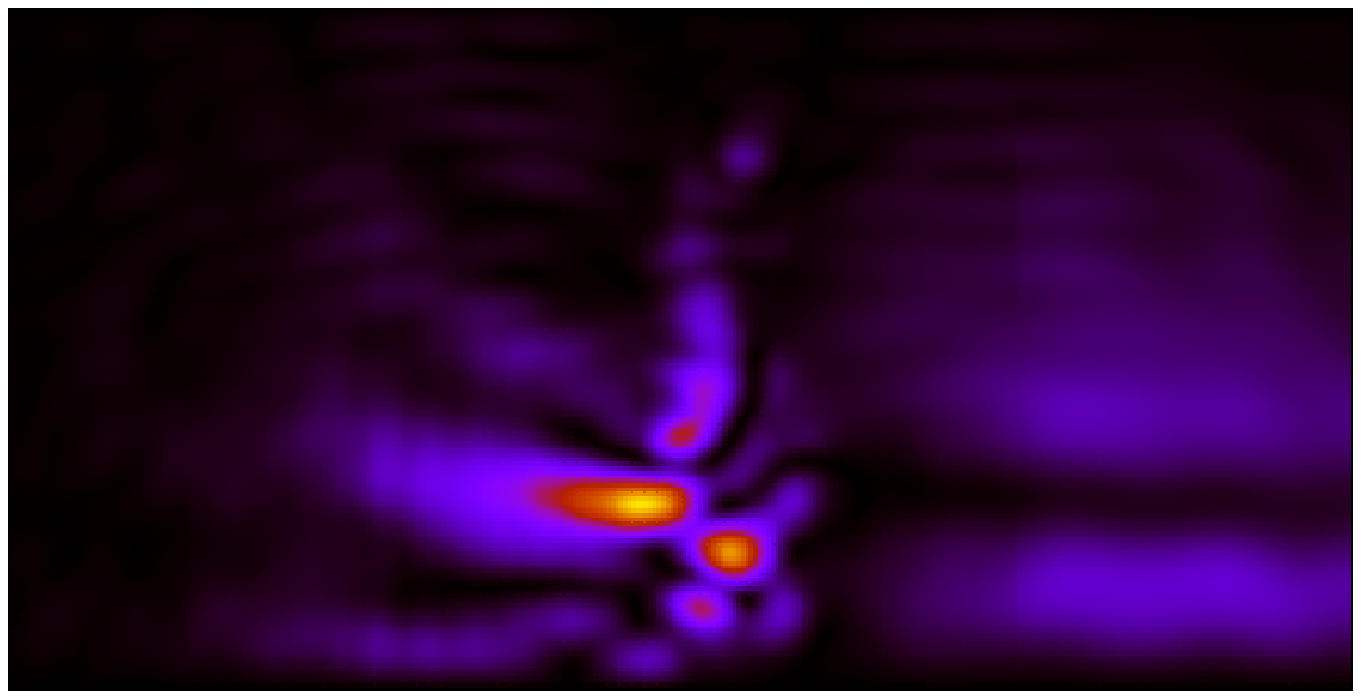} \\
\includegraphics[width=2.0cm]{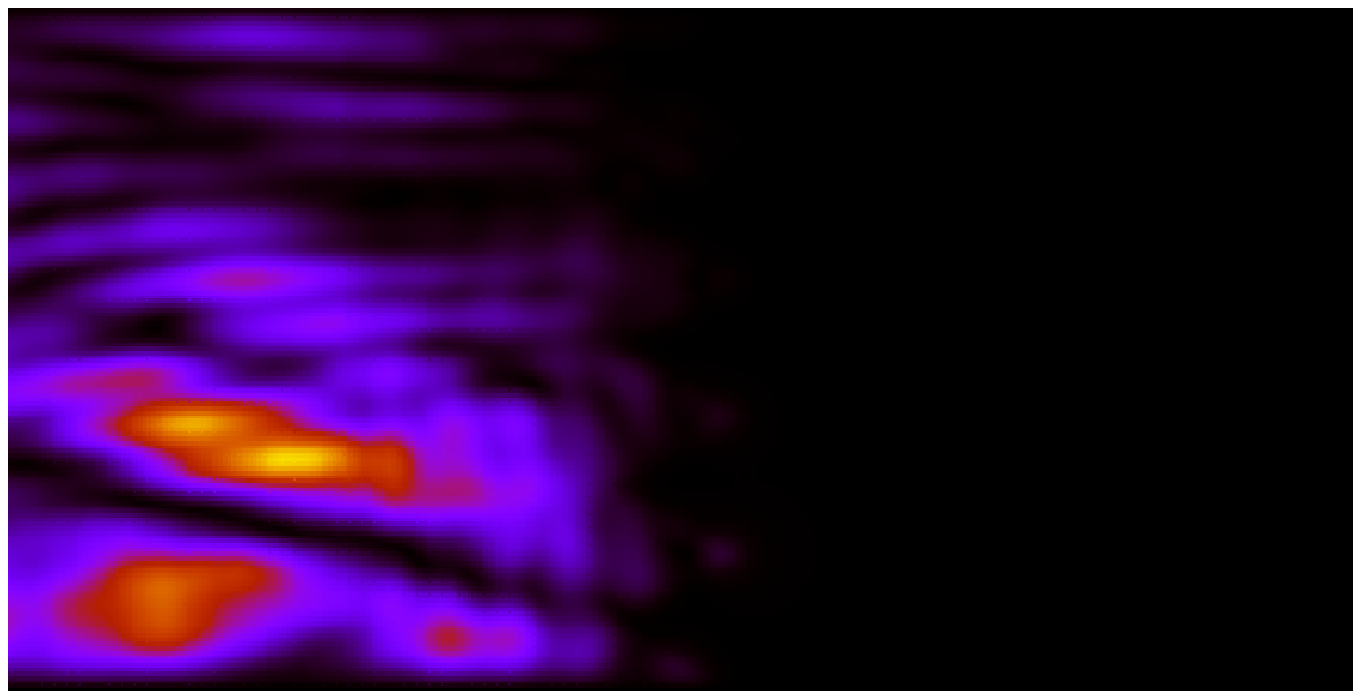}
\includegraphics[width=2.0cm]{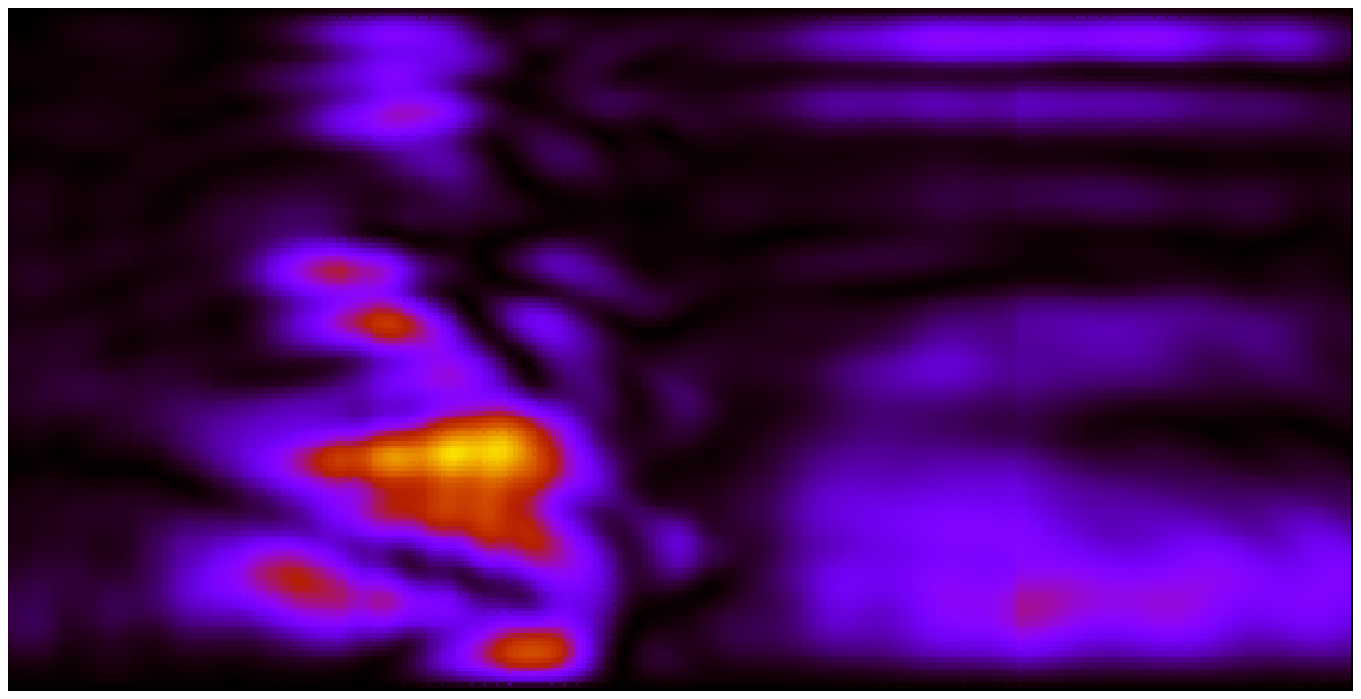} 
\includegraphics[width=2.0cm]{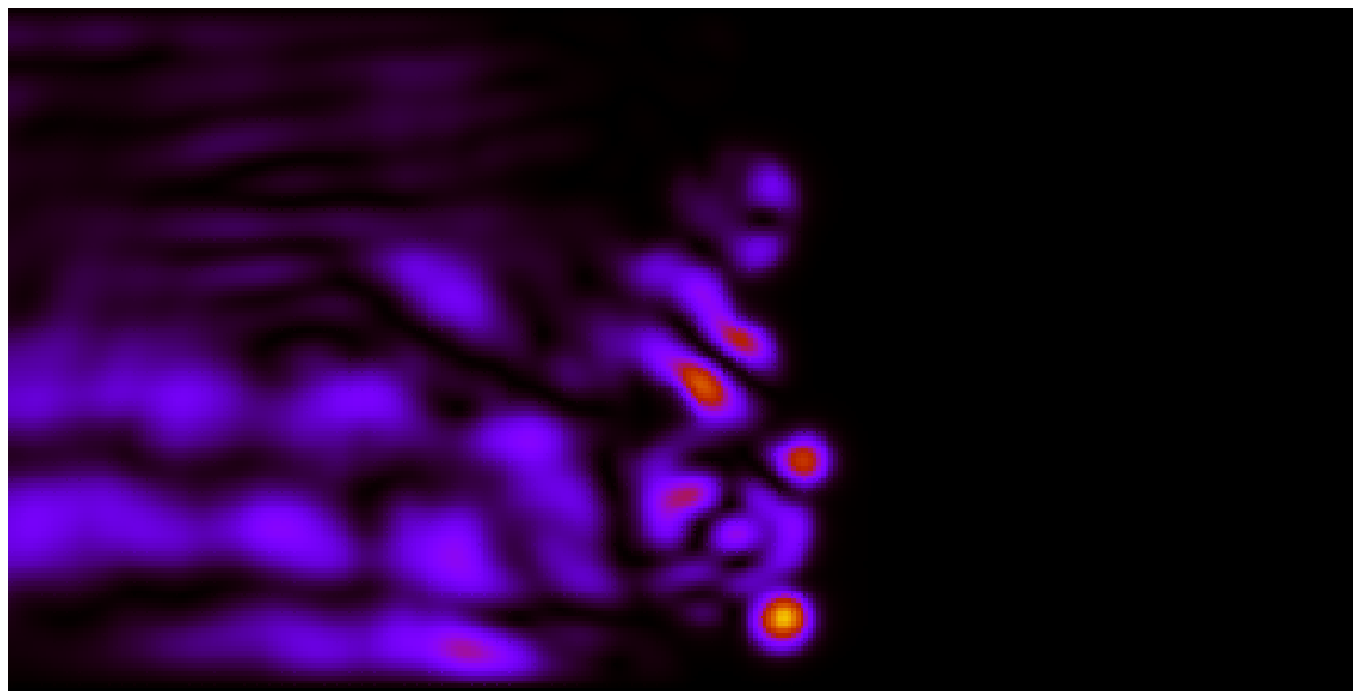}
\includegraphics[width=2.0cm]{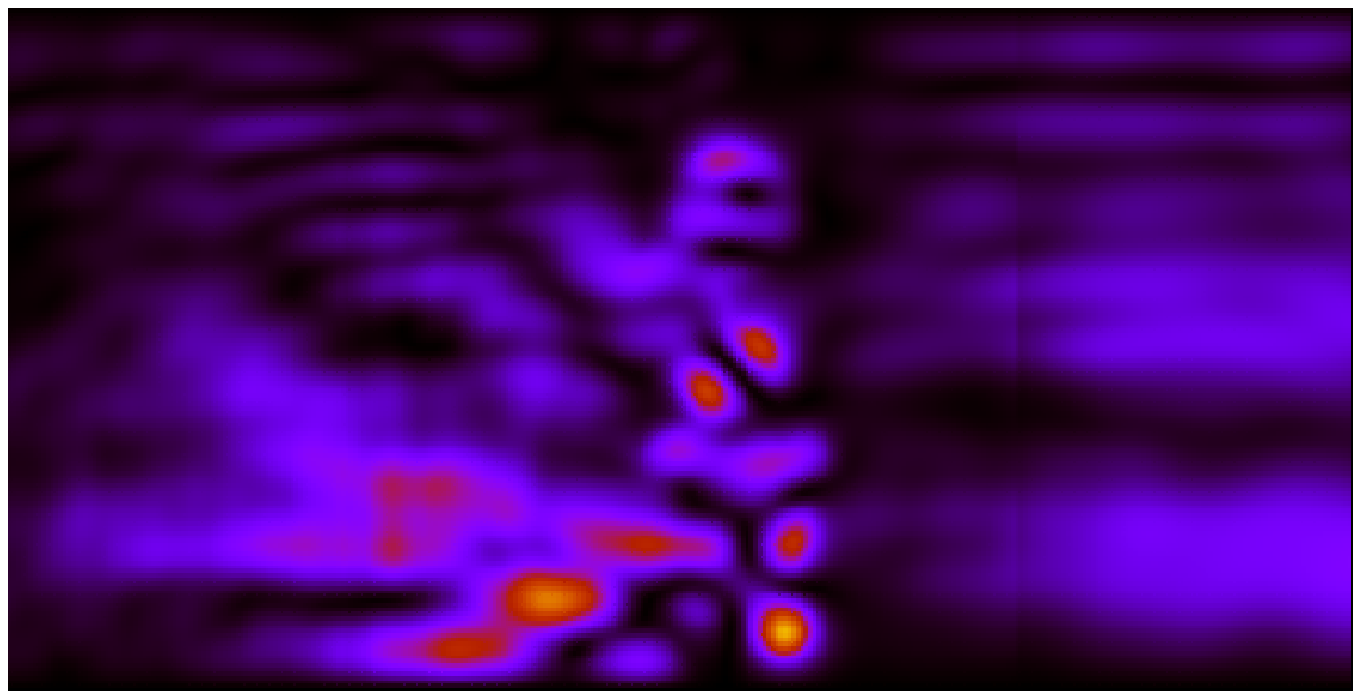} \\
\includegraphics[width=2.0cm]{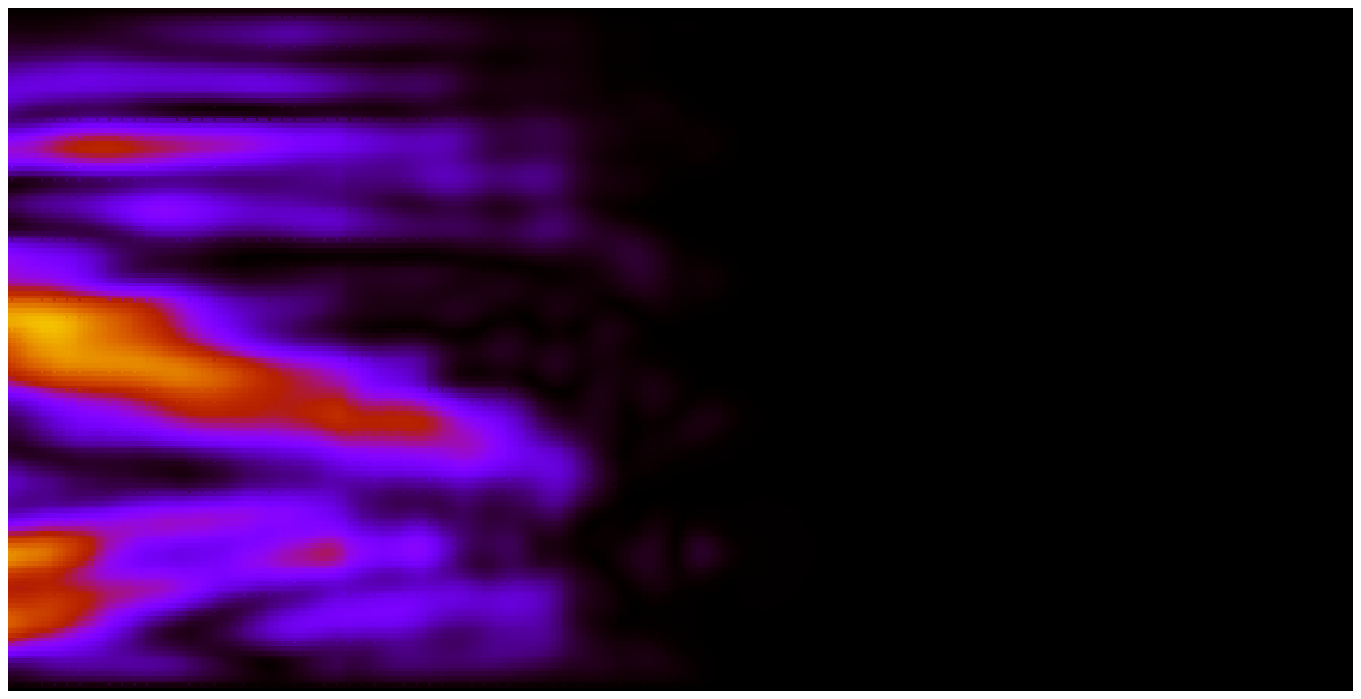}
\includegraphics[width=2.0cm]{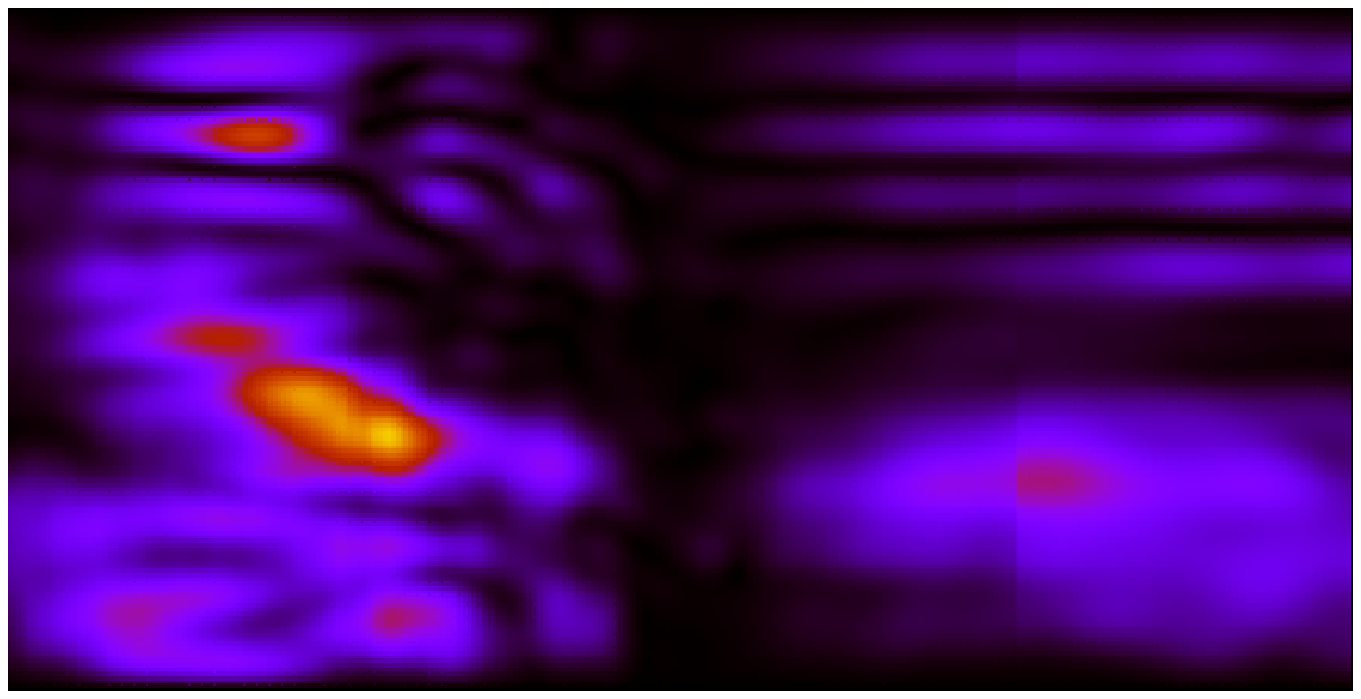} 
\includegraphics[width=2.0cm]{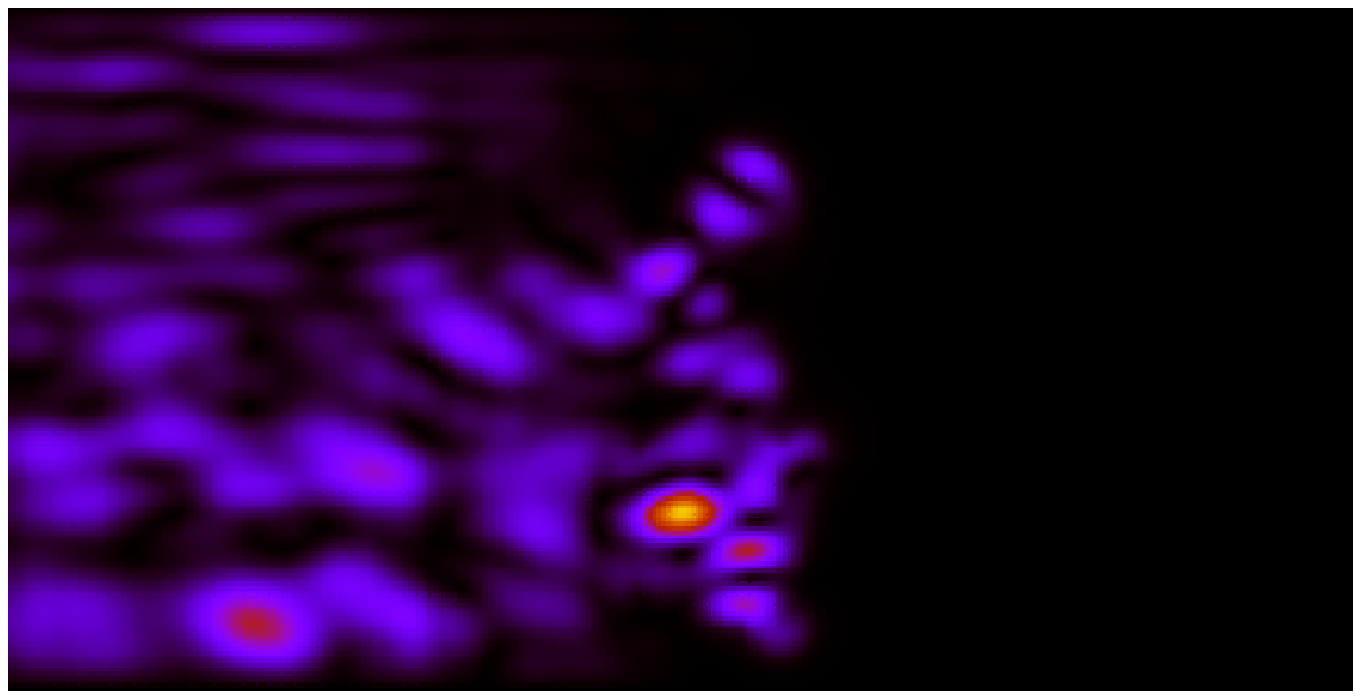}
\includegraphics[width=2.0cm]{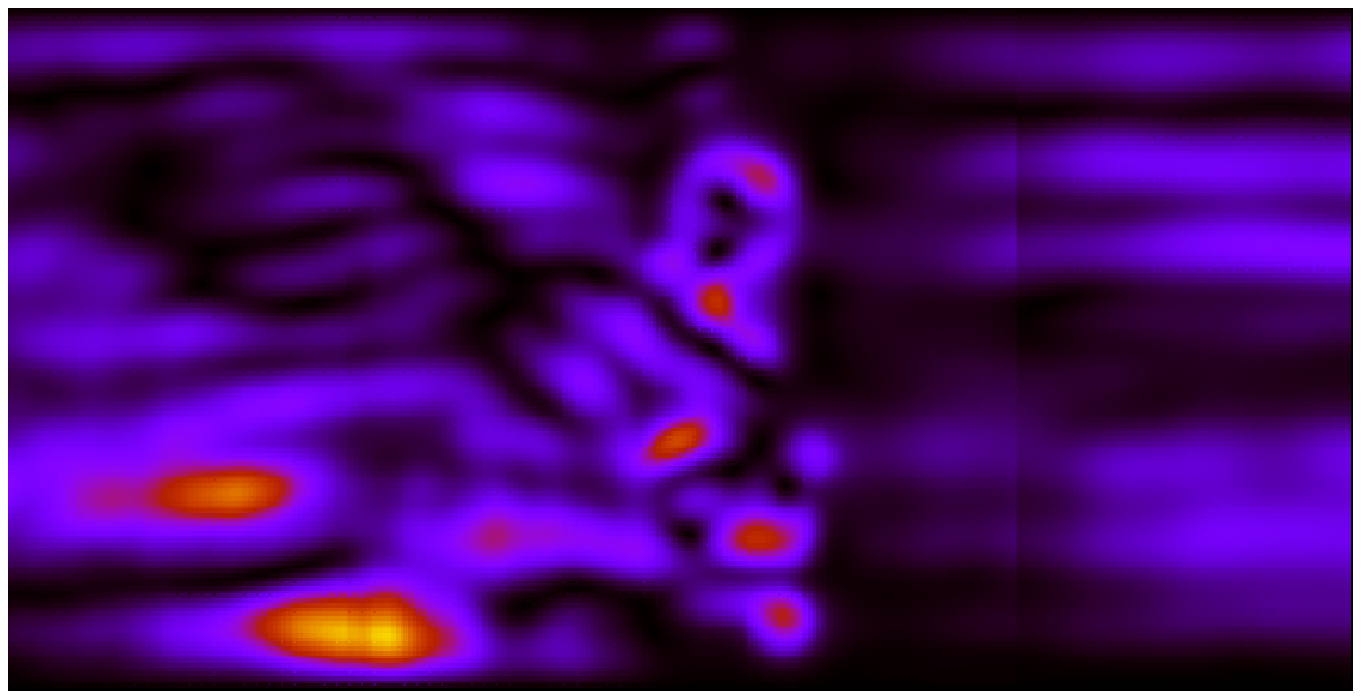} \\
\includegraphics[width=2.0cm]{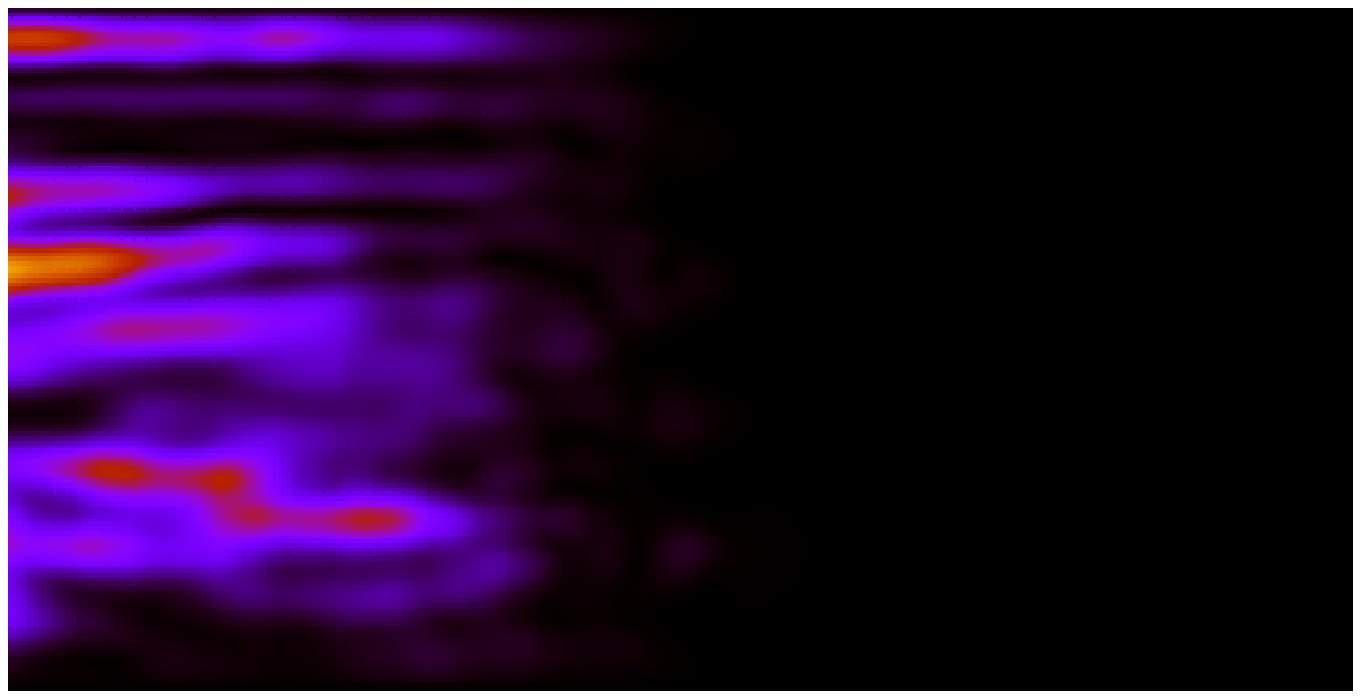}
\includegraphics[width=2.0cm]{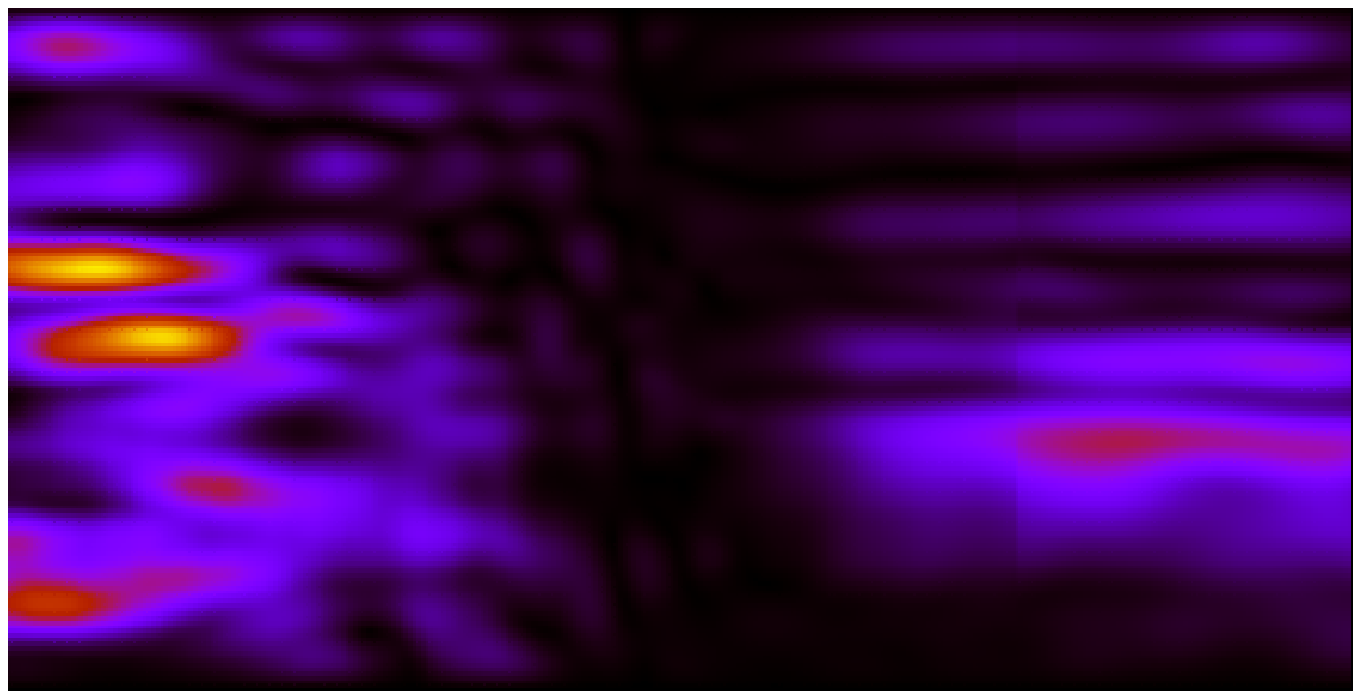} 
\includegraphics[width=2.0cm]{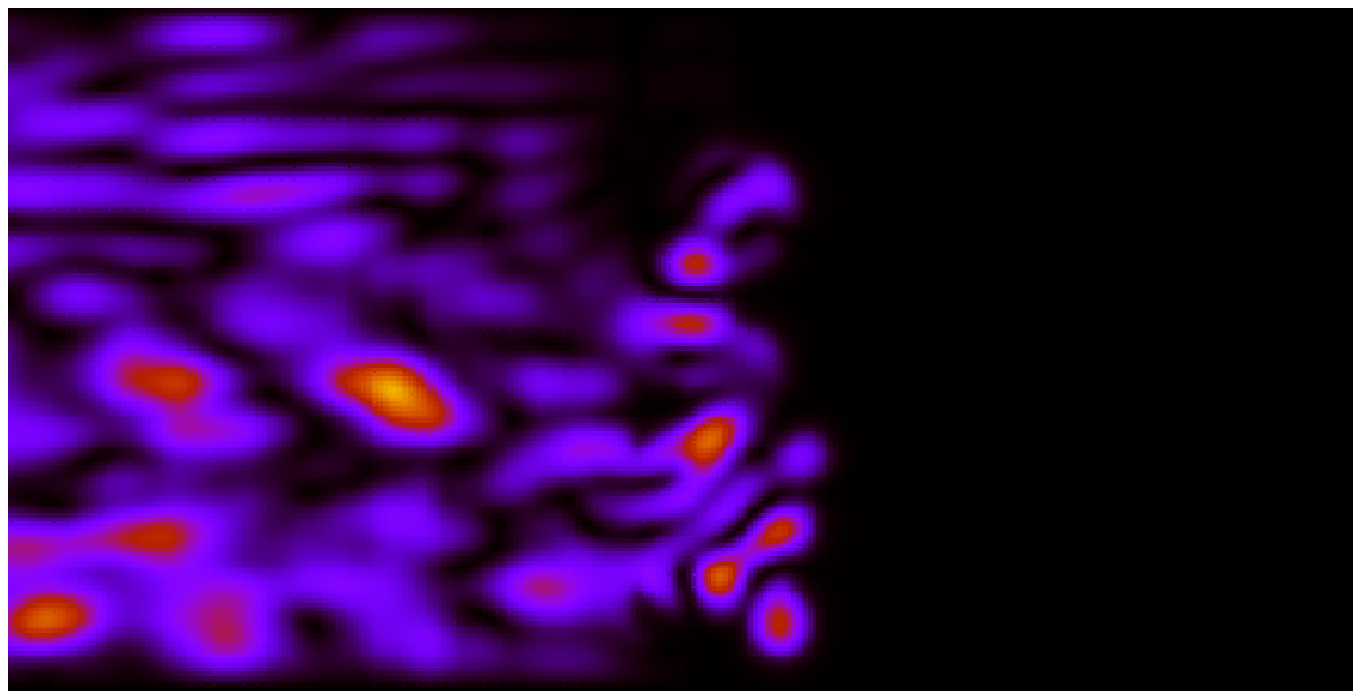}
\includegraphics[width=2.0cm]{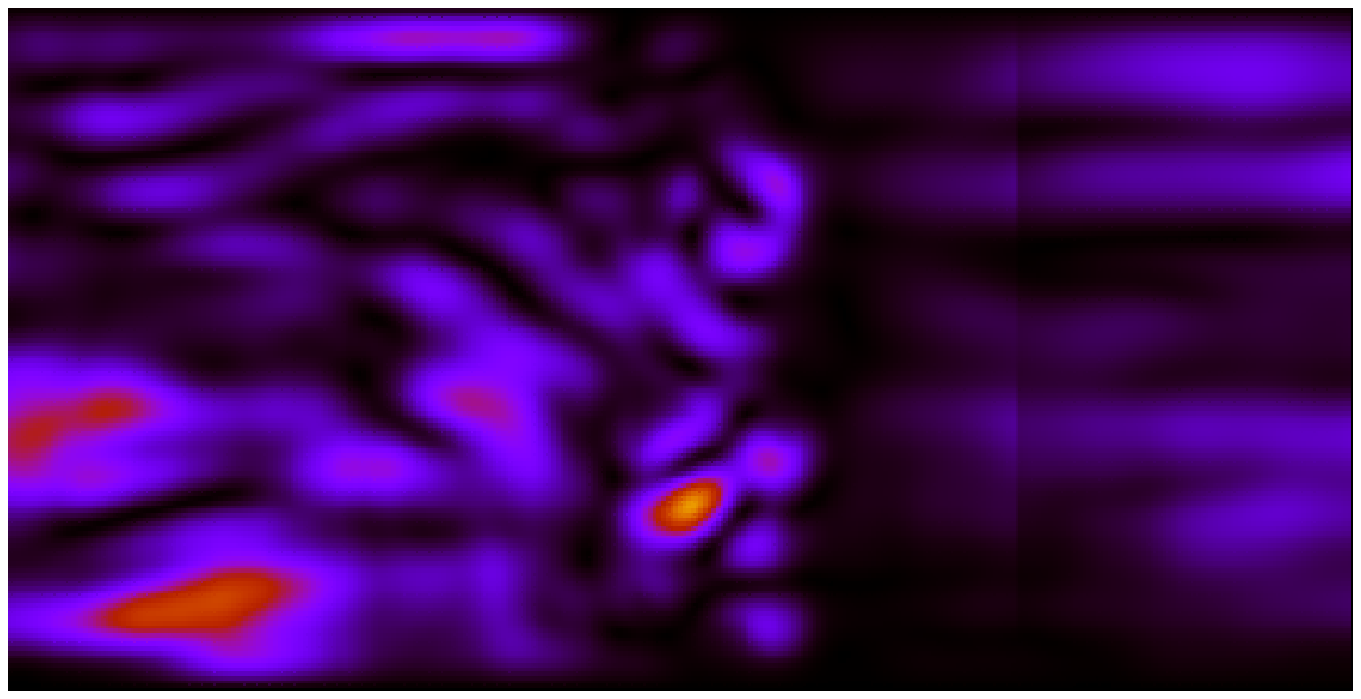} \\
\hspace*{0cm}(a)\hspace*{1.75cm}(b)\hspace*{1.75cm}(c)\hspace*{1.75cm}(d)
\caption{Snapshots of the wavepacket scattering for the zig-zag interface (a,b) and for the mesoporous interface (c,d), starting with $t_0=0$ in steps of $0.04$ ps from top to bottom. For (a,c) the bandoffset is $V_0=0.3$ eV, leading to a complete reflection of the wavepacket, while for (b,d) $V_0=0.1$ eV and partly transmitted electron waves are visible. For a better visibility, the colormap has been rescaled in each plot. 
}
\label{wps}
\end{figure}

We assume wavepackets are generated near the nanostructured interface, corresponding to either electrons or holes.
The initial wavepacket is described by a plane wave modulated by a Gaussian function 
\begin{equation}
\Psi_0(\rr) = A \cdot \Theta(\rr; \rr_0, R_c) \cdot \exp\left[-\frac{|\rr-\rr_{0}|^2}{2\sigma^2}\right] \cdot
			\exp(-i \bk_0 \rr),
\end{equation}
where the function $\Theta(\rr; \rr_0, R_c)$ is $1$ for $|\rr-\rr_0| \le R_c$ and $0$ for $|\rr-\rr_0| > R_c$,
 imposing strictly the vanishing condition for
$\Psi_0(\rr)$ beyond a cutting radius $R_c$ and $A$ is a normalization constant. 
By using the $\Theta$ function one can avoid introducing extra energy
in the wavepacket as it may partly overlap with high potential energy regions.

We consider the wavepacket is incident from the left, initially located at  
$\rr_0 = (x_0,y_0)$, where $x_0=-25$ nm, $y_0=-25$ nm for the zig-zag interface 
and $x_0=5$ nm, $y_0=-25$ nm for the mesoporous interface, i.e. it is placed 
in the middle of the largest 
circular nanoparticle. The wavevector $\bk_0\equiv(10\pi/L_x,0)$ was chosen along the $x$ direction, 
while $\sigma=L_x/16$ and the effective mass is $m^*=0.0655 m_0$ (GaAs).
The cutting radius is $R_c=10$ nm.
With these parameters the kinetic energy of the particle is $E_0=0.14$ eV.
A uniform grid with of 400 total energies $E_k$ in the range $[0,0.3]$ eV was considered.

Snapshots of wavepacket evolution are presented in Fig.\ \ref{wps}, for the 
zig-zag interface (a,b) and the mesoporous interface (c,d).
Two band offsets are considered, $V_0=0.1$ eV and $V_0=0.3$ eV, rendering a 
partially transparent interface and a practically opaque one, respectively.
The shapes of the interfaces introduce different consequences in the wavepacket scattering.
For example, the zig-zag interface tends to scatter significantly the wavepacket in the transversal direction, decreasing
the kinetic energy on the $x$ direction such that the propagation along the leads is slow.
By contrast, in the case of the mesoporous interface, the largest nanoparticle is the primary scatterer
and the scattered wave is more uniform.
In this case the partially transmitted and reflected waves are moving away from the interface at a 
higher speed compared to the zig-zag interface, but some probability remains confined for a longer time
at the mesoporous interface.   

Although the system considered here is two-dimensional, the method can be extended
to the more computationally demanding three-dimensional case.  
The qualitative picture regarding charge localization and transport is however qualitatively very similar, 
as it was pointed out in Refs. \cite{nemnes4,nemnes7}.

\subsection{Charge propagation near the nanostructured interface}

Based on our time-dependent approach with transparent boundary conditions we 
can calculate the evolution of the charge distribution and the photo-current. 
Wavepackets, which correspond to the electron/hole pairs generated at the interface, are scattered
elastically during a time interval which is a typical average coherence time $\tau_c$. 
Our aim here is not to fully describe the current collected at the contacts as it may depend on other
parameters, like e.g. recombination rates, but to describe the electron-hole separation at the 
nanostructured interface. 
Beyond $\tau_c$ the carriers may suffer inelastic scattering processes and, at even larger time scales, 
they diffuse towards the contacts moving through the bulk of the two materials, 
a process which may be further described by a classical drift-diffusion model.

\begin{figure}[t]
\centering
\includegraphics[width=6.cm]{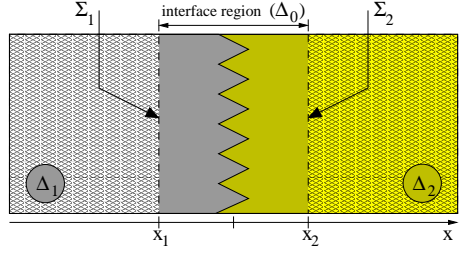}
\caption{Schematics of the photo-current model: the interface region $\Delta_0$ and the
outer regions $\Delta_1$ and $\Delta_2$.
\cblue{$\Sigma_s$ ($s$=1,2) represent planar interfaces $\Delta_s\cap\Delta_0$, located
at $x_s$.}
 Wavepackets are generated in $\Delta_0$ and 
the charge separation is evaluated for each region at a typical coherence time $\tau_c$.
}
\label{phc}
\end{figure}

In order to quantify the charge separation, 
we isolate the interface region $\Delta_0$ between positions $x_1$ and $x_2$. The outer regions 
$\Delta_1$ and $\Delta_2$, correspond to the two materials which make up the junction, as
depicted in Fig.\ \ref{phc}. These three domains may possibly correspond to the 
$\Omega_0$, $\Omega_s$ regions which previously defined the scattering problem.
We calculate the probability of finding the particle in the $\Delta_s$ domains, $Q_s$,
using the continuity relation for the probability current
%Assuming a given distribution of wavepacket parameters 
%($\bk_0$, $\sigma$, shape etc), we calculate the net charge $Q_s$ separated by the nanostructured interface
%using the continuity relation for the probability current
%and summing up the contributions for each wavepacket $w$:
\begin{equation}
Q_s(\tau_c) = \int_0^{\tau_c} dt \int_{\Sigma_s} {\bf j}(\rr,t) \; d{\bf \Sigma}_s 
            = \int_{\Delta_s} d\rr \; |\Psi(\rr,t=\tau_c)|^2, 
\end{equation}
where ${\bf j}(\rr,t) = 1/m \mbox{Re}[\Psi^*(\rr,t) (-i\hbar\nabla) \Psi(\rr,t)]$
is the probability current. The probabilities $Q_s$ correspond to the amounts
of charge separated by the interface, which are found in $\Delta_s$ domains at $\tau_c$.
Furthermore, one may define the current flowing into each $\Delta_s$ region as $I_s=d Q_s / dt$. 

\begin{figure}[t]
\centering
\includegraphics[width=8.cm]{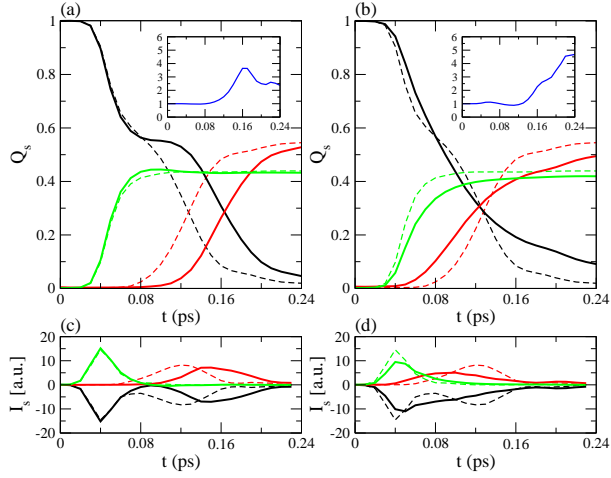}
\caption{Integrated probability flux $Q_s$ for the zig-zag (a) and mesoporous (b) interfaces.
$Q_s$ is calculated for the regions $\Delta0$ (black), $\Delta1$ (red), $\Delta2$ (green).
The dashed lines correspond to a flat interface. 
The insets show the ratio between $Q_0$ of zigzag/mesoporous interface and the same quantity
corresponding to the flat interface, as an indicator of trapped charge in the $\Delta_0$ region.
The photo-currents for each region $\Delta_s$ are indicated in (c) and (d).    
}
\label{iphc}
\end{figure}

The time evolution of the charge separation is plotted in Fig.\ \ref{iphc} for the zig-zag and mesoporous
interfaces for the cases shown in Fig.\ \ref{wps}(b) and \ref{wps}(d), 
in comparison with an ideally flat interface positioned at $x=0$. 
We define the regions $\Delta_s$ by choosing $x_1$ and $x_2$ according to Fig.\ \ref{phc} :
$x_1=-40$ nm, $x_2=10$ nm for the flat and zig-zag interfaces and 
$x_1=-10$ nm, $x_2=40$ nm for the mesoporous interface.
The interface regions have a length of 50 nm along the $x$ direction for all three structures and include 
the wave packet, which has the initial coordinates $(-25,-25)$ nm in the case of the flat and 
zig-zag interface and $(5,-25)$ nm in the case of the mesoporous interface.
The wave packet is therefore in both cases initially placed mid-way between $\Sigma_1$ and $\Sigma_2$.

In the following we discuss comparatively the particularities introduced by each type of interface. 
The chosen band offset of $V_0=0.1$ eV allows partly transmitted waves. 
For both nanostructured interfaces we obtain, in some respect, a similar behavior for 
the $Q_s$ time dependency. 
In the interface region $Q_0$ decreases, while $Q_1$ and $Q_2$ increase in time and tend to saturate.
The increase of $Q_1$ is delayed compared with $Q_2$ as the wave packet moves from left to right and
gets scattered at the interface.   
By comparison, keeping the same parameters, the zig-zag interface introduces larger delays 
compared with the flat one. This interface transfers part of the kinetic energy, initially on the 
$x$ direction, onto the $y$ direction, leading to a slower advance speed.
In contrast, the scattering across the mesoporous interface overall resembles better the data
obtained for the flat interface, with one difference: $Q_0$ decays slightly slower. 
The wave is trapped at the interface for longer times, 
as one can also see from the insets of Fig.\ \ref{iphc} 
and also from sequence in Figs.\ \ref{wps}(c) and (d),
compared with the zig-zag interface.
In addition, the increase of $Q_1$ and $Q_2$ is smoother compared to both flat and zig-zag interfaces.
The lower plots show the photo-currents in each region $\Delta_s$: the transmitted and reflected waves
are visible as current pulses of different widths. The back scattered waves arrive at the $\Delta_1$
region at later times compared to the transmitted waves and the width of the current pulse
is accordingly more dispersed.
Although the time-dependent quantities $Q_s(t)$ and $I_s(t)$ depend on the specific choice of the 
$\Delta_s$ domains, the observed behavior remains qualitatively the same.  

For a complete characterization of a certain interface one has to perform statistical averages, 
taking into account the variation of the wavepacket parameters ($\bk_0$, $\rr_0$, $\sigma$, shape etc).
Moreover, averages on different interfaces from the same class would be generally necessary.
To investigate the charge localization on either side of the interface, a distribution of $\tau_c$ may
also be considered and subsequently a drift-diffusion type model may be employed to
calculate the collected photo-current in a concrete device.
Here we presented the methodology for evaluating the charge separation
leaving a more complex analysis for a future study.

%\balance

\section{Conclusions}

We introduced a general framework for describing time-dependent coherent transport using the R-matrix
formalism. Expanding the time-dependent wavefunctions in the basis of scattering functions obtained
by solving the stationary problem for the open quantum system, 
the transparent boundary conditions are introduced in a natural way.   
The R-matrix method provides an efficient procedure of calculation the wavefunctions for a
relatively large energy set,
which is essential for an accurate description of the time-dependent transport.
The detailed steps of the computational scheme are provided. 
As applications, we consider the scattering of wavepackets across nanostructured interfaces,
which become increasingly relevant for new generation of nanostructured photovoltaic devices.
We discuss the transient behaviors of electrons crossing different types of interfaces.
Finally, we introduce a model for calculation of the photo-current and charge localization, which provides a tool for optimizing nanostructured interfaces. \\ 

{\bf Acknowledgment}\\

The research leading to these results has received funding from EEA
Financial Mechanism 2009 - 2014 under the project contract no 8SEE/30.06.2014 and
by the National Authority for Scientific Research and Innovation (ANCSI) under grant PN16420202.

\section*{References}

%\bibliography{mybibfile}

\end{document}